\documentclass{pasj00}
\draft

\begin{document}
\SetRunningHead{S. Kato}{Excitation of Disk Oscillations by Warps}
\Received{2004/5/5}
\Accepted{2004/7/2}

\title{Excitation of Disk Oscillations by Warps:  A Model of kHz QPOs}

\author{Shoji \textsc{Kato}%
  }
\affil{Department of Informatics, Nara Sangyo University, Ikoma-gun,
  Nara 636-8503}
\email{kato@io.nara-su.ac.jp; kato@kusastro.kyoto-u.ac.jp}


%


\maketitle

\begin{abstract}

The amplification of disk oscillations resulting from non-linear resonant
couplings between the oscillations and a warp is examined.
The disks are geometrically thin and general relativistic with 
non-rotating central objects.
By using a Lagrangian formulation, a general stability criterion is
derived.
The criterion is applied  to horizontal and vertical resonances of
g-mode oscillations and to horizontal resonances of p-mode
oscillations.
The results of analyses show that g-mode oscillations (including 
p-mode oscillations of $n=0$) are amplified by horizontal resonances with
the warp.
Other modes of oscillations with other resonances are all damped 
by resonances.
The amplified g-mode oscillations are located around the radius 
$4r_{\rm g}$, the radius of the epicyclic frequency being maximum, i.e., 
$\kappa_{\rm max}$.
The frequencies of amplified oscillations are harmonics of 
$\kappa_{\rm max}$, and well explain the 2 : 3 pairs of observed QPOs for 
reasonable masses (and spins) of the central objects.

\end{abstract}

\section{Introduction}


Kiloherz quasi-periodic oscillations (kHz QPOs) have been observed in
many X-ray binaries.
They are observed in both neutron stars and balck holes (for review,
e.g., van der Klis 2000).
Recently, they were also observed in X-ray pulsars (Wijnands et al. 2003).
One of the reasons why much attention is being paid to kHz QPOs
is that they give a clue for understanding the structure of the 
innermost region of relativistic disks, and provide powerful information
concerning the masses and spins of the central objects.

One of characteristics of kHz QPOs in black hole binaries is that they 
appear in pairs in many objects, and their frequency ratios are close 
to 2 : 3.
Paying attention on these points,
Abramowicz and Kluz{\' n}iak (2001) and Kluz{\' n}iak and Abramowicz (2001)
suggested that a parametric resonance is the cause of these QPOs.
Subsequently, they and their group developed their idea in many 
subsequent papers
(as recent papers, see, for example, Abramowicz, Klu{\'z}niak 2003;
Abramowicz et al. 2004a; Abramowicz et al. 2004b).
For a parametric resonance to work, disks must be deformed.
A well-known example of the excitation of oscillations on deformed disks is
a precession of disks deformed by a tidal force (Whitehurst 1988;
Hirose, Osaki 1990; Lubow 1991).
Another example is excitation of a spiral pattern in galactic disks
deformed by ram-pressure (Tosa 1994; Kato, Tosa 1994).
In the context of kHz QPOs, Kato (2003) considered disks deformed 
by a warp, and studied non-linear resonant interactions 
between the warp and g-mode oscillations on disks.
He showed that some g-mode oscillations are amplified on the disks by
resonances, and suggested that they would be the cause of QPOs.

The main purpose of this paper is to re-examine more generally
and more carefully the amplification of disk oscillations resulting from
resonant couplings between a warp and the oscillations,
correcting some errors in Kato (2003).
That is, we elaborate formulations while studying the condition of
nonlinear resonant amplifications by adopting the Lagrangian
formulation (Lynden-Bell, Ostriker 1967) of hydrodynamical equations. 
[In Kato (2003) Eulerian formulations are adopted.]
There are two superior points in Lagrangian formulations.
The first is that the non-linear coupling terms expressed in
Cartesian coordinates are compact.
The second is that the linear part of hydrodynamical equations can
be expressed in terms of Hermite operators (Lynden-Bell, Ostriker 1967).
This makes stability analyses perspective, and allows us to derive a
general stability criterion.

As a preparation for systematically studying which resonances amplify 
oscillations, the types of resonant couplings between 
a warp and oscillations were examined in detail in a previous paper 
(Kato 2004, hereafter Paper I).
The second purpose of this paper is to show, by calculating the growth
rates, which resonances really 
amplify oscillations, and to suggest the origin of the observed kHz QPOs.

\section{Basic Non-Linear Equations and a General Stability Criterion}

An unperturbed disk is geometrically thin, non-selfgravitating, and 
relativistic without accretion flow.
The effects of general relativity are taken into account within the framework 
of a pseudo-Newtonian formulation using the gravitational potential 
introduced by Paczy\'{n}ski and Wiita (1980), no rotation of the 
central object being assumed.
This means that all calculations of non-linear coupling terms are made 
within the framework of the Newtonian hydrodynamics, changing only the 
Newtonian gravitational potential to that of Paczy\'{n}ski and Wiita.

\subsection{Non-linear Hydrodynamical Equations of Lagrangian Formulation}

In a Lagrangian formulation, the hydrodynamical equation describing 
perturbations can be written as (Lynden-Bell, Ostriker 1967)
\begin{equation}
  {D_0^2\mbox{\boldmath $\xi$}\over Dt^2}=\delta\biggr(-\nabla\psi
                  -{1\over\rho}\nabla p\biggr),
\label{1.1}
\end{equation}
where \mbox{\boldmath $\xi$} is a displacement associated with perturbations, 
and $D_0/Dt$ is the time derivative along an unperturbed flow, 
$\mbox{\boldmath $u$}_0$,
and is related to the Eulerian time derivative, $\partial/\partial t$, by
\begin{equation}
  {D_0\over Dt}={\partial\over\partial t}+\mbox{\boldmath $u$}_0\cdot\nabla .
\label{1.2}
\end{equation}
In equation (\ref{1.1}), $\delta$ represents the Lagrangian variation
of the quantities in the subsequent parentheses, and $\psi$ is the 
gravitational potential.
Other notations in equation (\ref{1.1}) have their usual meanings.

The Lagrangian change of a quantity, $Q(\mbox{\boldmath $r$},t)$, i.e.,
$\delta Q$, is defined by
\begin{equation}
  \delta Q=Q[\mbox{\boldmath $r$}+\mbox{\boldmath $\xi$}
            (\mbox{\boldmath $r$},t)]
           -Q_0(\mbox{\boldmath $r$},t),
\label{1.3}
\end{equation}
where the subscript 0 to $Q$ represents the unperturbed value.
The Eulerian change of $Q$, i.e., $Q_1$, on the other hand, is defined by
\begin{equation}
    Q_1=Q(\mbox{\boldmath $r$},t)-Q_0(\mbox{\boldmath $r$},t).
\label{1.4}
\end{equation}
If $Q[\mbox{\boldmath $r$}+\mbox{\boldmath $\xi$}(\mbox{\boldmath $r$},t),t]$
is Taylor-expanded around $Q(\mbox{\boldmath $r$})$ up to the second-order 
terms with respect to perturbations, we have, from equations (\ref{1.3}) 
and (\ref{1.4}),
\begin{equation}
   \delta Q=Q_1+\xi_j{\partial Q_0\over\partial r_j}
               +\xi_j{\partial Q_1\over\partial r_j}
               +{1\over 2}\xi_i\xi_j{\partial^2 Q_0\over\partial 
                                                  r_i\partial r_j}.
\label{1.5}
\end{equation}
This is a relation between $\delta Q$ and $Q_1$, up to the second-order 
small quantities with respect to perturbations.
Here and hereafter, subscripts 0 and 1 are used to represent the unperturbed 
and (Eulerian) perturbed quantities, respectively.
Furthermore, here and hereafter, the summation abbreviation is adopted, 
using Cartescian coordinates.
(In the final calculations of the growth rate of the resonance interactions 
we adopt cylindrical coordinates.)

Our purpose here is to explicitly write down equation (\ref{1.1})
up to the second-order small quantities with respect to
perturbations.
Since the self-gravity of the disk is neglected here, i.e., $\psi_1=0$, 
we have easily
\begin{equation}
   \delta(\nabla\psi)=\xi_j{\partial\over\partial r_j}(\nabla\psi_0)
                      +{1\over 2}\xi_i\xi_j
               {\partial^2\over\partial r_i\partial r_j}(\nabla\psi_0).
\label{1.6}
\end{equation}
The second term on the right-hand side of equation (\ref{1.6}) represents
non-linear terms.
Expressing $\delta(\nabla p/\rho)$ in terms of Lagrangian quantities is,
on the other hand, somewhat complicated.
Using equation (\ref{1.5}) and the definition of Lagrangian change,
we have
\begin{eqnarray}
    \delta\biggr({1\over\rho}\nabla p\biggr)={1\over \rho_0+\delta\rho}
            \biggr[\nabla p_0+\nabla p_1+\xi_j{\partial\over\partial r_j}
                    \nabla(p_0+p_1)  \nonumber \\
            +{1\over 2}\xi_i\xi_j{\partial^2\over\partial r_i\partial r_j}
            \nabla p_0+...\biggr]-{1\over\rho_0}\nabla p_0.
\label{1.7}
\end{eqnarray}
The Eulerian pressure variation, $p_1$, which appears on the right-hand 
side of equation (\ref{1.7}), is expressed in terms of $\delta p$
and \mbox{\boldmath $\xi$} by using
\begin{equation}
  \delta p=p_1+\xi_j{\partial\over\partial r_j}(p_0+p_1)
          +{1\over 2}\xi_i\xi_j{\partial^2p_0\over\partial r_i\partial r_j}.
\label{1.8}
\end{equation}
We then have  
\begin{eqnarray}
    \rho_0\delta\biggr({1\over\rho}\nabla p\biggr)=\nabla\biggr[
          \delta p-\xi_j{\partial p_0\over\partial r_j}
          -\xi_j{\partial\over\partial r_j}\biggr(\delta p-
                    \xi_i{\partial p_0\over \partial r_i}\biggr)
          -{1\over 2}\xi_i\xi_j{\partial^2p_0\over\partial r_i\partial r_j} 
              \biggr]                    \nonumber           \\
          +\xi_j{\partial\over\partial r_j}\nabla\biggr
              [p_0+\delta p-\xi_i{\partial p_0\over \partial r_i}\biggr]
          +{1\over 2}\xi_i\xi_j{\partial^2\over\partial r_i\partial r_j}
                 \nabla p_0     \nonumber   \\
          -{\delta\rho\over \rho_0}\biggr[\nabla p_0+\nabla(\delta p)
              -(\nabla \xi_j){\partial p_0\over\partial r_j}\biggr]
          +\biggr({\delta\rho\over\rho_0}\biggr)^2\nabla p_0.
\label{1.9}
\end{eqnarray}
This explicitly expresses $\rho_0\delta(\nabla p/\rho)$ up to the 
second-order 
quantities in terms of Lagrangian qunatities, $\delta p$, 
$\delta\rho$, and $\mbox{\boldmath $\xi$}$.

Next, $\delta p$ and $\delta\rho$ on the right-hand side of equation
(\ref{1.9}) are explicitly expressed in terms of $\mbox{\boldmath $\xi$}$.
To do so we use the equation of continuity and the adiabatic
relation.
The former relation is expressed as (e.g., Kato, Unno 1967)
\begin{equation}
   \delta\rho+\rho_0{\partial\xi_i\over\partial r_i}=\rho_0N_{\rm c},
\label{1.10}
\end{equation}
where
\begin{equation}
   N_{\rm c}={1\over 2}\biggr[\biggr({\partial\xi_j\over\partial r_j}
                \biggr)^2             
        +{\partial\xi_i\over\partial r_j}{\partial \xi_j\over\partial r_i}
            \biggr].
\label{1.11}
\end{equation}
The adiabatic relation is written as (e.g., Kato, Unno 1967)
\begin{equation}
    {\delta p\over p_0}-\Gamma_1{\delta\rho\over\rho_0}=N_p,
\label{1.12}
\end{equation}
where
\begin{equation}
    N_p={1\over 2}\Gamma_1(\Gamma_1-1)
        \biggr({\delta\rho\over\rho_0}\biggr)^2.
\label{1.13}
\end{equation}

Substitution of equations (\ref{1.10}) and (\ref{1.12}) into 
equation (\ref{1.9}) finally gives an expression for 
$\rho_0\delta(\nabla p/\rho)$, expressed in terms of $\mbox{\boldmath
$\xi$}$ alone.
This is summarized as
\begin{equation}
       \rho_0\delta\biggr({1\over\rho}\nabla p\biggr)
             =\mbox{\boldmath $P$}_{(1)}
             +\mbox{\boldmath $P$}_{(2)},
\label{1.14}
\end{equation}
where $\mbox{\boldmath $P$}_{(1)}$ is the linear part with respect to 
$\mbox{\boldmath $\xi$}$ and
$\mbox{\boldmath $P$}_{(2)}$ is the non-linear part.
An explicit form of the linear part is given by Lynden-Bell and
Ostriker (1967), which is
\begin{equation}
    \mbox{\boldmath $P$}_{(1)}=\nabla\biggr[(1-\Gamma_1)p_0{\partial\xi_i
                      \over\partial r_i}\biggr]
                  -p_0\nabla\biggr({\partial\xi_i\over\partial r_i}\biggr)
                  -\nabla\biggr(\xi_j{\partial p_0\over\partial r_j}\biggr)
                  +\xi_j{\partial\over\partial r_j}\nabla p_0.
\label{1.15}
\end{equation}
The non-linear term, $\mbox{\boldmath $P$}_{(2)}$, generally has a 
complicated form.
In the case of isothermal disks, i.e., $\Gamma_1=1$, however, it has 
a simple expression.
We are satisfied here with using this simple expression for 
$\mbox{\boldmath $P$}_{(2)}$
in the following analyses of resonances, since the difference
of $\Gamma_1$ from unity will not bring about any qualitative differences
in the final results.
In the case of $\Gamma_1=1$, after some manipulations we have
\begin{equation}
    \mbox{\boldmath $P$}_{(2)}={\partial\over\partial r_i}
                  \biggr(p_0{\partial\xi_i\over\partial r_j}
                         \nabla\xi_j\biggr).
\label{1.16}
\end{equation}

Hereafter, we consider time-periodic perturbations, which vary as 
${\rm exp}(i\omega t)$, $\omega$ being a frequency.
Then, the non-linear hydrodynamical equation for adiabatic perturbations,
equation (\ref{1.1}), is written as
\begin{equation}
   -\omega^2\rho_0\mbox{\boldmath $\xi$}+2i\omega\rho_0
          (\mbox{\boldmath $u$}_0\cdot\nabla)\mbox{\boldmath $\xi$}
          +\mbox{\boldmath $L$}(\mbox{\boldmath $\xi$})
         =\rho_0\mbox{\boldmath $C$}
          (\mbox{\boldmath $\xi$},\mbox{\boldmath $\xi$}),
\label{1.17}
\end{equation}
where $\mbox{\boldmath $L$}$ is a linear operator with respect to 
\mbox{\boldmath $\xi$}, and is
(Lynden-Bell, Ostriker 1967)
\begin{equation}
    \mbox{\boldmath $L$}(\mbox{\boldmath $\xi$})
            =\rho_0(\mbox{\boldmath $u$}_0\cdot\nabla)
              (\mbox{\boldmath $u$}_0\cdot\nabla)\mbox{\boldmath $\xi$}
            +\rho_0(\mbox{\boldmath $\xi$}\cdot\nabla)(\nabla\psi_0)
            +\mbox{\boldmath $P$}_{(1)}
               (\mbox{\boldmath $\xi$}).
\label{1.18}
\end{equation}
The non-linear term $\rho_0$\mbox{\boldmath $C$} consists of two parts, i.e., 
$\rho_0\mbox{\boldmath $C$}=\rho_0\mbox{\boldmath $C$}_\psi
+\rho_0\mbox{\boldmath $C$}_p$,
where $\rho_0\mbox{\boldmath $C$}_\psi$ comes from 
$-\rho_0\delta(\nabla\psi)$ and
$\rho_0\mbox{\boldmath $C$}_p$ from $-\rho_0\delta(\nabla p/\rho)$.
From equation (\ref{1.6}) we have
\begin{equation}
   \rho_0\mbox{\boldmath $C$}_\psi=-{1\over 2}\rho_0\xi_i\xi_j
       {\partial^2\over\partial r_i\partial r_j}(\nabla\psi_0),
\label{1.19}
\end{equation}
and from equation ({\ref{1.16}) we have
\begin{equation}
    \rho_0\mbox{\boldmath $C$}_p=-{\partial\over\partial r_i}
         \biggr(p_0{\partial\xi_i\over\partial r_j}\nabla\xi_j\biggr).
\label{1.20}
\end{equation}
Detailed expressions for $\rho_0\mbox{\boldmath $C$}_\psi$ and
$\rho_0\mbox{\boldmath $C$}_p$ in cylindrical coordinates are given in
appendix 1.

An important characteristics of the linear operator $\mbox{\boldmath $L$}$ 
is that it is a Hermitian (Lynden-Bell, Ostriker 1967).
This characteristics of \mbox{\boldmath $L$} is used later when we evaluate 
the growth rate of oscillations.

\begin{figure}
  \begin{center}
    \FigureFile(80mm,80mm){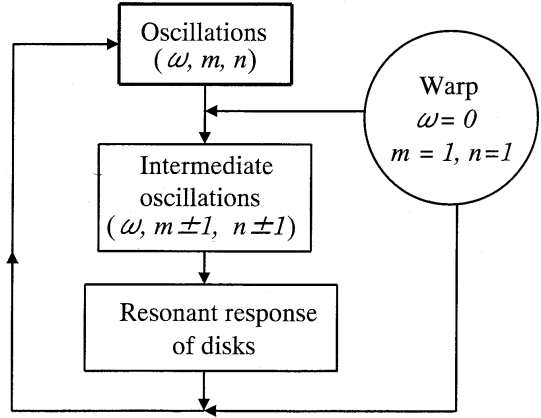}
  \end{center}
  \caption{
Feedback processes of nonlinear resonant interactions acting on
oscillations.
The original oscillations are characterized by $\omega$, $m$, and $n$.
Since the warp corresponds to a wave mode of $\omega=0$, $m=1$, and $n=1$,
the nonlinear interaction between them brings about 
intermediate modes of oscillations of $\omega$, $m\pm 1$, and $n\pm 1$.
To these intermediate oscillations, the disk resonantly responds at
a certain radius.
Then, the intermediate oscillations feedback to the original oscillations
after the resonance.
This feedback process amplifies or dampens
the original oscillations, since resonances are involved in the interaction 
processes. (After Paper I)
  }
  \label{fig:1}
\end{figure}

\subsection{A General Method for Deriving Stability Criterion}

We assume that a mode of oscillations is present on the disk
with a small amplitude.
The displacement associated with the oscillations is denoted by
$\mbox{\boldmath $\xi$}$, which satisfies the homogeneous part of
the wave equation (\ref{1.17}).
Now, we further assume that the unperturbed disk is deformed by 
some external forces from an equilibrium state.
The displacement associated with the deformations is denoted by
$\mbox{\boldmath $\xi$}^{\rm w}$. 
Nonlinear couplings between $\mbox{\boldmath $\xi$}$ and
$\mbox{\boldmath $\xi$}^{\rm w}$ are now examined, assuming that
they are weak and their effects can be studied by
applying successive approximations.
The processes of the couplings are shown schematically in figure 1.
The first step of nonlinear couplings between $\mbox{\boldmath $\xi$}$ and
$\mbox{\boldmath $\xi$}^{\rm w}$ is their direct interactions.
The resulting oscillations are denoted by $\mbox{\boldmath $\xi$}^{\rm int}$, 
and hereafter called intermediate oscillations.
Since we are interested here in the case when the time scale associated 
with the deformation is sufficiently long compared with $1/\omega$, 
the intermediate oscillations are described by using the inhomogeneous
wave equation (\ref{1.17}) as 
\begin{equation}
   -\omega^2\rho_0\mbox{\boldmath $\xi$}^{\rm int}+2i\omega\rho_0
          (\mbox{\boldmath $u$}_0\cdot\nabla)\mbox{\boldmath $\xi$}^{\rm int}
          +\mbox{\boldmath $L$}(\mbox{\boldmath $\xi$}^{\rm int})
         =\rho_0\mbox{\boldmath $C$}
          (\mbox{\boldmath $\xi$},\mbox{\boldmath $\xi$}^{\rm w}).
\label{1.17'}
\end{equation}

As the next step of nonlinear interactions, we consider interactions
between the intermediate oscillations $\mbox{\boldmath $\xi$}^{\rm int}$ and 
the deformations $\mbox{\boldmath $\xi$}^{\rm w}$.
This interaction is interesting, since it generally feedbacks to the 
original oscillations of $\mbox{\boldmath $\xi$}$.    
Unless a resonance phenomenon is involved in the feedback processes, 
the effects of the feedbacks are only to change the frequencies and
functional forms of the original oscillations. 
If a resonance process is involved, however, there is a possibility of
the oscillations being amplified (or damped).
The Lagrangian formulation of hydrodynamical equations by Lynden-Bell 
and Ostriker (1967) makes this analysis easy.
The equation describing these feedbacks is given by
\begin{equation}
   -\omega^2\rho_0\mbox{\boldmath $\xi$}+2i\omega\rho_0
          (\mbox{\boldmath $u$}_0\cdot\nabla)\mbox{\boldmath $\xi$}
          +\mbox{\boldmath $L$}(\mbox{\boldmath $\xi$})
         =\rho_0\mbox{\boldmath $C$}
          (\mbox{\boldmath $\xi$}^{\rm int},\mbox{\boldmath $\xi$}^{\rm w}).
\label{1.17''}
\end{equation}
That is, the left-hand side of the equation is the same as equation
(\ref{1.17}), while the coupling term is
$\rho_0\mbox{\boldmath $C$}
   (\mbox{\boldmath $\xi$}^{\rm int},\mbox{\boldmath $\xi$}^{\rm w})$.
Let us integrate the equation over a volume with relevant
boundary conditions after multiplying by $\mbox{\boldmath $\xi$}^*$,
where the asterisk denotes the complex conjugate.

Since the operators on the left-hand side of equation (\ref{1.17''}) are
Hermitians, the integrations
\begin{equation}
   i \int \rho_0\mbox{\boldmath $\xi$}^*(\mbox{\boldmath $u$}_0\cdot
          \nabla)\mbox{\boldmath $\xi$}dV
\qquad {\rm and} \qquad 
   \int \mbox{\boldmath $\xi$}^* \mbox{\boldmath $L$}(\mbox{\boldmath $\xi$})dV
\end{equation}
are real.
The integration of the right-hand side of the equation, on the
other hand, has an imaginary part when resonances are involved in the 
interactions.
This implies that $\omega$ must be complex. 
If we write $\omega=\omega_{\rm r}+i\omega_{\rm i}$, the imaginary part of 
equation (\ref{1.17''}) becomes
\begin{equation}
   -2i\omega_{\rm i}\int\mbox{\boldmath $\xi$}^{*}
        \rho_0[\omega-i(\mbox{\boldmath $u$}_0\cdot\nabla)]
        \mbox{\boldmath $\xi$}dV
      = \Im \int \rho_0 \mbox{\boldmath $\xi$}^* 
            \mbox{\boldmath $C$}(\mbox{\boldmath $\xi$}^{\rm int}, 
        \mbox{\boldmath $\xi$}^{\rm w})dV.
\label{1.17'''}
\end{equation}
Hereafter, $\omega_{\rm r}$ is denoted by $\omega$ without any confusion.
From this equation we can calculate $\omega_{\rm i}$.
In the subsequent sections we examine this issue for the cases where the 
disk deformations are due to warps.

Equation (\ref{1.17'''}) has a clear physical meaning.
The integral on the left-hand side is the wave energy, $E$, times $2/\omega$
[see equation (92) of Kato (2001)].
Hence, the left-hand side of equation (\ref{1.17'''}) represents the 
growing rate of the wave energy, $-2i\omega_{\rm i}E$, times $2/\omega$.
This implies that the right-hand side of equation (\ref{1.17'''})
must be the work, $W$, done on the original oscillations by the force
$\mbox{\boldmath $C$}(\mbox{\boldmath $\xi$}^{\rm int},
\mbox{\boldmath $\xi$}^{\rm w})$ during the time interval $2/\omega$.
In other words, the work done on the oscillations by resonances, $W$, 
during a unit time is 
\begin{equation}
     W={\omega\over 2}\Im \int \mbox{\boldmath $\xi$}^*
            \mbox{\boldmath $C$}(\mbox{\boldmath $\xi$}^{\rm int},
            \mbox{\boldmath $\xi$}^{\rm w})dV.
\label{1.18}
\end{equation}
This is understandable, since if the velocity associated with the 
oscillations and the force, $\mbox{\boldmath $C$}$, are in phase 
as an idealized situation, the work done on
the oscillations during which the displacement becomes 
$\mbox{\boldmath $\xi$}$ (the quarter of one cycle) is 
$
        \Im \int \mbox{\boldmath $\xi$}^*
            \mbox{\boldmath $C$}(\mbox{\boldmath $\xi$}^{\rm int},
            \mbox{\boldmath $\xi$}^{\rm w})dV.
$
Hence, the rate of work (per unit time) is, on average, 
$
        \Im \int \mbox{\boldmath $\xi$}^*
            \mbox{\boldmath $C$}(\mbox{\boldmath $\xi$}^{\rm int},
            \mbox{\boldmath $\xi$}^{\rm w})dV
$
times $2\omega/\pi$.
It is finally noted that the wave enegy, $E$, is expressed later in a 
more conventional form.

\section{Warps, Oscillations, and Dispersion Relation}

The unperturbed disk is assumed to have no motion except for rotation.
The rotation is further assumed to be cylindrical.
That is, with cylindrical coordinates ($r, \varphi, z$) whose $z$-axis 
is perpendicular to 
the disk plane and whose origin is at the disk center, the angular velocity 
of rotation, $\Omega$, is represented as $\Omega=\Omega(r)$.
When we need numerical figures, we adopt the Keplerian form for $\Omega$, 
i.e., $\Omega=\Omega_{\rm K}(r)$, since geometrically thin disks 
are considered here.

The unperturbed disk is further assumed to be isothermal in the vertical 
direction.
In vertically isothermal disks the density is stratified as 
(e.g., Kato et al. 1998)
\begin{equation}
     \rho_0(r,z)=\rho_{00}(r){\rm exp}\biggr(-{z^2\over 2H^2(r)}\biggr),
\label{1.21}
\end{equation}
where $\rho_{00}$ is the density on the equatorial plane and the
half-thickness of the disk, $H$, is related to the angular velocity of the
Keplerian rotation, $\Omega_{\rm K}$, by
\begin{equation}
     \Omega_{\rm K}^2(r)H^2(r)={p_0\over\rho_0}=c_{\rm s}^2,
\label{1.22}
\end{equation}
where $c_{\rm s}$ is the isothermal speed of sound.

\subsection{Oscillations Modes and Warps}

Oscillations on disks are generally a combination of motions on the 
equatorial plane and those in the vertical direction.
These motions are coupled in complicated ways, and
cannot be treated separately.
In the case of the vertically isothermal disks, however, we can
approximately separate them (Okazaki el al. 1987; see also Kato et al. 1998).
In order to make analyses simple, we introduce here an approximation.
In this case, the $z$-dependence of the eigen-functions of disk oscillations 
are described by Hermite polynomials, ${\cal H}_n(z/H)$, $n$ specifying the
node number in the vertical direction.
The horizontal displacements associated with the
oscillations, $\mbox{\boldmath $\xi$}_\bot^{(0)}$, are then described by
\begin{equation}
    \mbox{\boldmath $\xi$}_\bot^{(0)}={\cal H}_n
                {\rm exp}[i(\omega t-m\varphi)]
                \hat {\mbox{\boldmath $\xi$}}_\bot^{(0)}(r)
\label{1.23}
\end{equation}
and that in the vertical direction, $\xi_z^{(0)}$, is
\begin{equation}
    \xi_z^{(0)}={\cal H}_{n-1}
                {\rm exp}[i(\omega t-m\varphi)]
                {\hat {\xi}}_z^{(0)}(r).
\label{1.23'}
\end{equation}
Here, $\mbox{\boldmath $\xi$}_\bot$ is an abbreviation of the horizontal
displacement vector 
$(\xi_r, \xi_\varphi)$ and $m$ is the number of arms in the azimuthal
direction.
The oscillations are now classified by a set of $m$ and $n$.
Furthermore, for a set of ($m$, $n$), we have two kinds of oscillations, 
which are p-mode oscillations and g-mode ones.
The formers are oscillations of $(\omega-m\Omega)^2>n\Omega$
and the latters are those of $(\omega-m\Omega)^2<\kappa^2$ 
(see e.g., Kato et al. 1998; Kato 2001).
The p-mode oscillations start from $n=0$, while the g-mode ones start from
$n=1$.
In the framework of the present approximations, the p-mode oscillations
of $n=0$ are purely horizontal.
In order to make it clear, the coefficient $n$ has been attached to
equation (\ref{1.23'}).

In addition to the oscillations mentioend above, the unperturbed disk 
is assumed to be deformed by a global standing $(\omega=0)$ pattern 
of a warp, $\mbox{\boldmath $\xi$}^{\rm w}$,  
which can be approximated in the lowest approximation to be (e.g., Kato 2003)
\begin{equation}
    \xi_r^{\rm w}=\xi_\varphi^{\rm w}=0
\label{1.25}
\end{equation}
and
\begin{equation}
     \xi_z^{\rm w}={\cal H}_0{\rm exp}(-i\varphi)\eta(r).
\label{1.26}
\end{equation}
The present warp corresponds to a radially long wavelength perturbation 
with $n=1$ and $m=1$.
Here and hereafter, the superscript w is used to represent the 
quantities associated with the warp.

\subsection{Dispersion Relation}

As a preparation to solve the inhomogeneous wave equation (\ref{1.17}),
we discuss here the homogeneous part of the equation.
The homogeneous part is generally a set of partial differential
equations.
In some cases, however, the homogeneous equation becomes a set of
algebraic equations, and thus the inhomogeneous non-linear equation 
(\ref{1.17}) becomes manageable.
As such a case, we discuss the case where the wavelength of
perturbations is sufficiently short compared with the radius of the
disk.
This approximation is allowed, except for some special situations, since 
we consider geometrically thin disks.

Here, we assume that ${\mbox{\boldmath $\xi$}} \propto {\rm exp}(ikr)$ and
$kr\gg 1$.
Since the z-dependences of ${\mbox {\boldmath $\xi$}}$ are
given by equations (\ref{1.23}) and (\ref{1.23'}), using 
equations (\ref{1.15}) and (\ref{1.18}) and 
the characteristics of the Hermite polynomials 
we have, after some manupulations, that the homogeneous part of
equation (\ref{1.17}) is expressed as
\begin{equation}
  [-(\omega-m\Omega)^2+\kappa^2-4\Omega^2]{\hat \xi}_r^{(0)}
   -i2\Omega(\omega-m\Omega){\hat \xi}_\varphi^{(0)}
   +(kH)\Omega_\bot^2\biggr[(kH)\hat{\xi}_r^{(0)}+i\hat{\xi}_z^{(0)}\biggr]
      =0,
\label{1.26a}
\end{equation}
\begin{equation}
  -(\omega-m\Omega)^2{\hat \xi}_\varphi^{(0)}+i2\Omega(\omega-m\Omega)
   {\hat \xi_r}^{(0)}
   =0,
\label{1.26b}
\end{equation}
and
\begin{equation}
   [-(\omega-m\Omega)^2+n\Omega_\bot^2]{\hat \xi}_z^{(0)}
    -in(kH)\Omega_\bot^2{\hat \xi}_r^{(0)}=0.
\label{1.26c}
\end{equation}
In these equations the terms with $c_{\rm s}^2$ come from the Eulerian
pressure variations (see the next paragraph).
If these terms are neglected, the above equations are those derived by
the particle approximations, and thus the horizontal motions and the 
vertical motions are uncoupled.
This is also shown clearly by deriving the dispersion relation (i.e.,
the solvability condition) from the set of equations 
(\ref{1.26a})--(\ref{1.26c}), which is
\begin{equation}
  [(\omega-m\Omega)^2-\kappa^2][(\omega-m\Omega)^2-n\Omega_\bot^2]
      =c_{\rm s}^2k^2(\omega-m\Omega)^2.
\label{disp}
\end{equation}

It will be instructive to note here that for a polytropic gas
we can derive after manipulations that ${\mbox {\boldmath $P$}_1}({\mbox
{\boldmath $\xi$}})$ is related to $\rho_0\nabla h_1$ by
\begin{equation}
  \mbox{\boldmath $P$}_{(1)}(\mbox{\boldmath $\xi$})=\rho_0\nabla h_1+\rho_0
  (\mbox{\boldmath $\xi$}\cdot\nabla)
              \biggr({1\over\rho_0}\nabla p_0\biggr),
\label{2.3}
\end{equation}
where $h_1\equiv p_1/\rho_0$ is the Eulerian pressure variation, $p_1$,
being normalized by $\rho_0$.
In the framework of the present approximations, 
$\partial h_1/\partial r=c_{\rm
s}^2[k^2{\hat \xi}_r^{(0)}+i(k/H){\hat \xi}_z^{(0)}]{\cal H}_n$ and 
$\partial h_1/\partial z= [n\Omega_\bot^2{\hat \xi}_z^{(0)}-inc_{\rm
s}^2(k/H){\hat \xi}_r^{(0)}]{\cal H}_{n-1}$.

\section{The First Stage of Coupling between Warp and Oscillations}

We here examine how the disk oscillations and the warp described in 
section 3 are coupled.
The procedures of examinations are the same as those in Paper I
and outlined in subsection 2.2.
That is, we assume that the amplitude of oscillations is 
so small that their non-linear effects on the oscillations, themselves, 
are negligible.
The amplitude of the warp, on the other hand, is so large that the effects 
of the warp on the oscillations are non-negligible.
We examine them successively, as is shown in figure 1.
The first step of their interactions is direct non-linear coupling between 
the warp and the oscillations, which brings about intermediate
oscillations of ($m\pm 1$, $n\pm 1$) with frequency $\omega$
(see figure 1). 
To these intermediate oscillations the disk responds resonantly at a
radius where the dispersion relation to the intermediate oscillations is 
satisfied.
After having this resonant interaction, the intermediate oscillations
interact with the warp  to feedback to the original oscillations of
($m$, $n$) with $\omega$ (see figure 1).
This non-linear feedback to the original oscillations through a
resonant coupling amplifies or dampens the original oscillations.

In this section we consider the first stage of the above couplings.
The intermediate waves resulting from this coupling are expressed by attaching 
a superscript (1) as $\mbox{\boldmath $\xi$}^{(1)}$.
As mentioned before, the coupling has two paths.
One is the path where the azimuthal wavenumber of the intermediate 
oscillations is $m+1$, and the other is the path with $m-1$.
To distinguish $\mbox{\boldmath $\xi$}^{(1)}$ in these two paths, 
we attach the 
subscript $+$ to $\mbox{\boldmath $\xi$}^{(1)}$ as 
$\mbox{\boldmath $\xi$}_+^{(1)}$ in the former
case and the subscript $-$ in the latter case as ${\bf \xi}_-^{(1)}$.
From equation (\ref{1.17}) the inhomogeneous wave equation describing 
$\mbox{\boldmath $\xi$}_\pm^{(1)}$ is found to be
\begin{equation}
  -\omega^2\rho_0\mbox{\boldmath $\xi$}_\pm^{(1)}
        +2i\omega\rho_0(\mbox{\boldmath $u$}_0\cdot\nabla)
            \mbox{\boldmath $\xi$}_\pm^{(1)}
        +\mbox{\boldmath $L$}(\mbox{\boldmath $\xi$}_\pm^{(1)})
        = \rho_0\mbox{\boldmath $C$}_\pm^{(1)},
\label{2.1}
\end{equation}
where
\begin{equation}
    \mbox{\boldmath $C$}_+^{(1)}=\mbox{\boldmath $C$}
    (\mbox{\boldmath $\xi$}^{(0)},\mbox{\boldmath $\xi$}^{\rm w})
    \qquad{\rm and}\qquad
    \mbox{\boldmath $C$}_-^{(1)}=\mbox{\boldmath $C$}
    (\mbox{\boldmath $\xi$}^{(0)},\mbox{\boldmath $\xi$}^{{\rm w}*}).
\label{2.2}
\end{equation}
Equation (\ref{2.1}) represents two equations.
One is the equation for the upper sign (it is $+$ in the present case) and
the other is that for the lower sign (e.g., $-$).
Both equations are written in one set, since they have similar forms
and can be treated in parallel.

Our purpose here is to solve this inhomogeneous wave equation (\ref{2.1})
to obtain $\mbox{\boldmath $\xi$}_+^{(1)}$ and 
$\mbox{\boldmath $\xi$}_-^{(1)}$.
Although this equation can be formally solved by using a Green function, 
it is too complicated and not practical.
Hence, we are satisfied here to solve the equation
by introducing the approximations of vertically isothermal disks
(see section 3).
In this case, concerning the functional forms in the vertical direction,
the feedbacks to the original oscillations occur through two paths.
One is a path through $n$ $\rightarrow$ $n-1$ $\rightarrow$ $n$
(lower path).
The other is through the path of $n$ $\rightarrow$ $n+1$
$\rightarrow$ $n$ (upper path).
In the former case, the intermediate oscillations in the
horizontal direction, $\mbox{\boldmath $\xi$}_\bot^{(1)}$, are 
proportional to ${\cal H}_{n-1}(z/H)$, while it is
proportional to ${\cal H}_{n+1}(z/H)$ in the latter case.
[It is noted that the vertical displacemtnt, $\xi_z^{(1)}$, associated 
with the intermediate oscillations are 
proportional to ${\cal H}_{n-2}(z/H)$ and ${\cal H}_{n}(z/H)$.] 
In order to treat these two cases separately, we attach the subscript
${\tilde n}$ to $\mbox{\boldmath $\xi$}_\pm^{(1)}$ as
$\mbox{\boldmath $\xi$}_{\pm,{\tilde n}}^{(1)}$, where ${\tilde n}$
represents $n-1$ or $n+1$.
Then, the $r$-, $\varphi$-, and $z$-components of 
$\mbox{\boldmath $\xi$}_{\pm,{\tilde n}}^{(1)}$ are written as
\begin{equation}
   \xi_{r,\pm,{\tilde n}}^{(1)}
     ={\cal H}_{\tilde n}{\rm exp}\{i[\omega t-(m\pm 1)\varphi]\}
      \hat{\xi}_{r,\pm, {\tilde n}}^{(1)}(r),
\label{2.13a}
\end{equation}
\begin{equation}
   \xi_{\varphi,\pm,{\tilde n}}^{(1)}
     ={\cal H}_{\tilde n}{\rm exp}\{i[\omega t-(m\pm 1)\varphi]\}
      \hat {\xi}_{\varphi,\pm, {\tilde n}}^{(1)}(r),
\label{2.13b}
\end{equation}
\begin{equation}
   \xi_{z,\pm,{\tilde n}}^{(1)}
     ={\cal H}_{{\tilde n}-1}{\rm exp}\{i[\omega t-(m\pm 1)\varphi]\}
      \hat {\xi}_{z,\pm,{\tilde n}}^{(1)}(r).
\label{2.13c}
\end{equation}
It should be noted here that in equation (\ref{2.13c}) subscript
${\tilde n}$ is attached to $\xi_{z,\pm}^{(1)}$, although it is proportional
to ${\cal H}_{{\tilde n}-1}$.
We adopt this notation in order to clarify that 
$\xi_{r,\pm, {\tilde n}}^{(1)}$,
$\xi_{\varphi,\pm,{\tilde n}}^{(1)}$, and $\xi_{z,\pm,{\tilde n}}^{(1)}$
are one of the sets of solutions for the same intermediate oscillations.

Using equations (\ref{2.13a})--(\ref{2.13c}) 
and the procedures used to derive the wave equations for
oscillations $\mbox{\boldmath $\xi$}^{(0)}$ 
[equations (\ref{1.26a})--(\ref{1.26c})], 
we can write the equations corresponding to equation (\ref{1.17'}) as
\begin{eqnarray}
    \{-[\omega-(m\pm1)\Omega]^2&&+(\kappa^2-4\Omega^2)\}
    {\hat \xi}_{r,\pm,{\tilde n}}^{(1)}
   -i2\Omega[\omega-(m\pm 1)\Omega]{\hat \xi}_{\varphi,\pm,{\tilde n}}^{(1)}
       \nonumber \\
    & & +k^2c_{\rm s}^2{\hat \xi}^{(1)}_{r,\pm,{\tilde n}}
    +ik{c_{\rm s}^2\over H}{\hat \xi}^{(1)}_{z,\pm,{\tilde n}}
       ={\hat C}_{r,\pm,{\tilde n}}^{(1)},
\label{2.7}
\end{eqnarray}
\begin{equation}
   -[\omega-(m\pm 1)\Omega]^2{\hat \xi}_{\varphi,\pm,{\tilde n}}^{(1)}
   +i2\Omega[\omega-(m\pm 1)\Omega]{\hat \xi}_{r,\pm,{\tilde n}}^{(1)}
       = {\hat C}_{\varphi,\pm,{\tilde n}}^{(1)},
\label{2.8}
\end{equation}
\begin{equation}
   \{-[\omega-(m\pm 1)\Omega]^2
        +{\tilde n}\Omega_\bot^2\}{\hat \xi}_{z,\pm,{\tilde n}}^{(1)}
   -ikn{c_{\rm s}^2 \over H}{\hat \xi}^{(1)}_{r,\pm,{\tilde n}}    
     = {\hat C}_{z,\pm,{\tilde n}}^{(1)},
\label{2.9}
\end{equation}
where ${\hat C}_{r,\pm,{\tilde n}}^{(1)}$ and ${\hat C}_{\varphi,\pm,{\tilde
n}}^{(1)}$ are the parts of $C_{r,\pm,{\tilde n}}^{(1)}$ and
$C_{\varphi,\pm,{\tilde n}}^{(1)}$ that are proportional to
${\rm exp}\{i(\omega t-(m\pm 1)\varphi)\}{\cal H}_{\tilde n}(z/H)$,
respectively, i.e.,
\begin{equation}
   C_{r,\pm,{\tilde n}}^{(1)}={\cal H}_{\tilde n}(z/H)
            {\rm exp}\{i(\omega t-(m\pm 1)\varphi)\}
            {\hat C}_{r,\pm,{\tilde n}}^{(1)},
\end{equation}
\begin{equation}
   C_{\varphi,\pm,{\tilde n}}^{(1)}={\cal H}_{\tilde n}(z/H)
            {\rm exp}\{i(\omega t-(m\pm 1)\varphi)\}
            {\hat C}_{\varphi,\pm,{\tilde n}}^{(1)}.
\end{equation}
Similarly, 
\begin{equation}
   C_{z,\pm,{\tilde n}}^{(1)}={\cal H}_{{\tilde n}-1}(z/H)
           {\rm exp}\{i(\omega t-(m\pm 1)\varphi)\}
           {\hat C}_{z,\pm,{\tilde n}}^{(1)}.
\end{equation}
Detailed expressions for ${\hat C}_{r,\pm,{\tilde n}}^{(1)}$,
${\hat C}_{\varphi,\pm,{\tilde n}}^{(1)}$, and 
${\hat C}_{z,\pm,{\tilde n}}^{(1)}$ are given in appendix 2.

Equation (\ref{2.7})--(\ref{2.9}) are easily solved with respect to 
$\mbox{\boldmath ${\hat \xi}$}^{(1)}$ to obtain
\begin{equation}
    {\hat \xi}_{r,\pm,{\tilde n}}^{(1)}
        ={1\over D_{\pm, {\tilde n}}}{\hat \zeta}_{r,\pm,{\hat n}}^{(1)},
   \quad
   {\hat \xi}_{\varphi,\pm,{\tilde n}}^{(1)}
           ={1\over D_{\pm, {\tilde n}}}{\hat \zeta}_{\varphi,\pm,{\hat n}}
            ^{(1)},
   \quad {\rm and} \quad
    {\hat \xi}_{z,\pm,{\tilde n}}^{(1)}={1\over D_{\pm,{\tilde n}}}
           {\hat \zeta}_{z,\pm,{\hat n}}^{(1)},
\end{equation}
where
\begin{eqnarray}
    \hat{\zeta}_{r,\pm,{\tilde n}}^{(1)}=
        & &-[(\omega-(m\pm 1)\Omega)^2-{\tilde n}\Omega_\bot^2]
         \biggr[{\hat C}_{r,\pm,{\tilde n}}^{(1)}
       -i{2\Omega\over \omega-(m\pm 1)\Omega}{\hat C}_{\varphi,\pm,{\tilde n}}
          ^{(1)}\biggr]   \nonumber \\
        & &-i(kH)\Omega_\bot^2{\hat C}_{z,\pm,{\tilde n}}^{(1)},
\label{2.10}
\end{eqnarray}
\begin{equation}
   \hat{\zeta}_{\varphi, \pm,{\tilde n}}^{(1)}=
    i {2\Omega\over \omega-(m\pm 1)\Omega}\hat{\zeta}_{r,\pm,{\tilde n}}^{(1)},
\label{2.11}
\end{equation}
\begin{equation}
    \hat{\zeta}_{z,\pm, {\tilde n}}^{(1)}
   =i{\tilde n}\Omega_\bot^2
   \biggr\{kH\biggr[{\hat C}_{r,\pm,{\tilde n}}^{(1)}
       -i{2\Omega\over \omega-(m\pm 1)\Omega}{\hat C}_{\varphi,\pm,{\tilde n}}
          ^{(1)}\biggr]
   +i{[\omega-(m\pm 1)\Omega]^2-\kappa^2 \over
      [\omega-(m\pm 1)\Omega]^2}   
      {\hat C}_{z,\pm,{\tilde n}}^{(1)}\biggr\}.
\label{2.12}            
\end{equation}
Furthermore, $D_{\pm,{\tilde n}}$ is defined as
\begin{equation}
    D_{\pm, {\tilde n}} =\biggr\{[\omega-(m\pm 1)\Omega]^2-\kappa^2\biggr\}
       \biggr\{[\omega-(m\pm 1)\Omega]^2-{\tilde n}\Omega_\bot^2\biggr\}
      -k^2c_{\rm s}^2[\omega-(m\pm 1)\Omega]^2.
\label{2.13}
\end{equation}

An important characteristics of the intermediate mode, 
$\mbox{\boldmath $\xi$}^{(1)}$,
is that it becomes infinite at the radius where $D_{\pm, {\tilde n}} =0$.
This implies that the disk responds resonantly to the intermediate 
oscillations at these radii.
Since the pressure terms in equation (\ref{2.13}) are usually small, 
$D_{\pm, {\tilde n}} =0$ is realized at radii close to 
$[\omega-(m\pm 1)\Omega]^2-\kappa^2 \sim 0$ or 
$[\omega-(m\pm 1)\Omega]^2-{\tilde n}\Omega_\bot^2\sim 0$.
The former resonances are hereafter called
horizontal resonances, while the latter ones are vertical resonances.

\section{Resonant Radii and Resonant Frequencies}

The resonant radii where the disk responds resonantly to the intermediate 
oscillations must be within the 
propagation region of the original oscillations.
This condition has been extensively examined by Paper I.
We mention here a part of the results.

There are two kinds of oscillations on disks, i.e.,
the p-mode and g-mode oscillations.
Furthermore, resonances have two types, i.e., horizontal and 
vertical resonances.
Combinations of them give us three types of resonances:
i) horizontal resonances of the g-mode oscillations, ii) vertical
resonances of the g-mode oscillations, and iii)
horizontal resonances of the p-mode oscillations.
It should be noted that vertical resonances of the p-mode oscillations 
are absent.

As a demonstration, we present here where horizontal resonances of 
the g-mode oscillations occur and how much their frequencies are when 
$n=1$ and the path of the coupling is $n$ $\rightarrow$ $n+1$ and $n$ 
(i.e., ${\tilde n}=2$). 
As equation (\ref{2.13}) shows, the horizontal resonances occur near to
\begin{equation}
  \omega-(m\pm 1)\Omega\sim  \kappa \qquad {\rm or}\qquad
  \omega-(m\pm 1)\Omega\sim -\kappa.
\label{3.01}
\end{equation}
The propagation of the g-mode oscillations with frequency $\omega$,
on the other hand, is bound in the region of 
$-\kappa<\omega-m\Omega<\kappa$ [see dispersion
relation (\ref{disp})].
Since we are interested in oscillations whose frequencies are near to
the boundaries of the propagation region, we require that
\begin{equation}
    \omega-m\Omega\sim -\kappa \qquad {\rm or} \qquad
    \omega-m\Omega\sim \kappa.
\label{3.02}
\end{equation}
Requirements (\ref{3.01}) and (\ref{3.02}) are simultaneously satisfied
when $\Omega\sim 2\kappa$.
That is, in the case of the plus sign of $m\pm 1$, they are satisfied 
when $\omega-(m+1)\Omega\sim -\kappa$ and $\omega-m\Omega
\sim \kappa$, and in the case of the minus sign of $m\pm 1$,
$\omega-(m-1)\Omega\sim \kappa$ and $\omega-m\Omega\sim
-\kappa$.
The requirement $\Omega=2\kappa$  
is realized at $r\sim 4.0r_g$ when the Schwarzschild metric is adopted,
where $r_g$ is the Schwarzschild radius, 
defined by $r_g=2GM/c^2$, $M$ being the mass of the central object.
The frequencies of the oscillations at the radius are $m\Omega\mp \kappa$.
In the case of $m=2$ they are 0.1875 and 0.313 in units of 
$(GM/r_g^3)^{1/2}$ (see Paper I).

The above discussions are only for the case of the horizontal resonances of
the g-mode oscillations with $n=1$ and ${\tilde n}=2$.
Discussions for other types of resonances (Paper I) show that
resonances can occur at, and only at, the radii where
\begin{equation}
   \kappa=0, \quad  \sqrt{2}\Omega_\bot-\Omega=\kappa,
             \quad  \Omega=2\kappa, \quad 
   \sqrt{3}\Omega_\bot-\Omega=\kappa,
\label{3.03}
\end{equation}
depending on the types of the resonances and modes of oscillations.
The radii satisfying the above requirements in equation (\ref{3.03})
are, in turn,
\begin{equation}
   r=3r_g, \quad r=3.62r_g, \quad r=4.0r_g, \quad {\rm and}\quad
   r=6.46r_g.
\label{3.04}
\end{equation}
Since more detailed discussions concerning the resonant radii and the 
resonant frequencies at these radii are given by Paper I,
we proceed here to the issue which resonances lead 
to amplifications of the oscillations.

To proceed further, however, there is an important issue to be noted
here, which is the sign of the ratio of $\omega-(m\pm 1)\Omega$ 
and $\omega-m\Omega$ at the resonance point.
This is important to discuss whether the resonances lead to 
amplification or to damping, as will become clear in section 7.
The sign is negative in the case demonstrated above.
Careful examinations show that the ratio is always negative in 
horizontal resonances of all g-mode oscillations, while
it is positive in vertical resonances of all g-mode oscillations
and horizontal resonances of p-mode oscillations (except for $n=0$).

\section{The Second Stage of Coupling}

The next issue is to study whether the resonances discussed in the 
previous sections act so as to amplify the original oscillations or not.
For this purpose, the non-linear coupling between the intermediate 
oscillations 
$\mbox{\boldmath $\xi$}^{(1)}$ and the warp 
$\mbox{\boldmath $\xi$}^{\rm w}$ is discussed.
The coupling brings about oscillations of the same form as the 
original oscillations, i.e.,
the $r$- and $\varphi$- components of the displacements of the resulting 
oscillations are proportional to  
exp$[i(\omega t-m\varphi)]{\cal H}_n(z/H)$ and 
the $z$-component is proportional to 
exp$[i(\omega t-m\varphi)]{\cal H}_{n-1}(z/H)$.
This feedback acts 
so as to amplify or dampen the  original oscillations, depending on 
the phase relations at the resonance points.

Now, we start from equation (\ref{1.17'''}).
The equation has a clear physical meaning.
The left-hand side of the equation can be reduced to
\begin{equation}
   -2i\omega_{\rm i}\int(\omega-m\Omega)\rho_0
      (\xi^{(0)*}_r\xi^{(0)}_r
      +\xi^{(0)*}_z\xi^{(0)}_z)dV,
\label{4.4}
\end{equation}
when an approximate relation between $\xi_{r}^{(0)}$ and $\xi_\varphi^{(0)}$,
i.e.,
\begin{equation}
    i(\omega-m\Omega)\xi^{(0)}_\varphi+2\Omega\xi^{(0)}_r=0,
\label{4.5}
\end{equation}
which is the $\varphi$-component of the equation of motion, is used.
If vertical integration is performed, equation (\ref{4.4}) can be 
written as 
\begin{equation}
    -2i\omega_{\rm i}(2\pi)^{1/2}H
     \int (\omega-m\Omega)\rho_{00}
     [n\hat {\xi}_r^{(0)*}\hat {\xi}_r^{(0)}+
      (n-1)\hat {\xi}_z^{(0)*}\hat {\xi}_z^{(0)}]2\pi rdr.
\label{4.4'}
\end{equation}
We now remember that the wave energy is known to be defined by (e.g., 
Kato 2001)
\begin{equation}
    E={1\over 2}\omega\int (\omega-m\Omega)\rho_0
        (\xi_r^{(0)*}\xi_r^{(0)}+\xi_z^{(0)*}\xi_z^{(0)})dV.
\label{4.4''}
\end{equation}
[In Kato (2001) $E$ is written by using the Eulerian velocity perturbations,
$u_r$ and $u_z$, instead of the Lagrangian displacements.]
Hence, equation (\ref{4.4}) is 
\begin{equation}
        -4i{\omega_{\rm i}\over \omega_0}E,
\label{4.4left}
\end{equation}
and is related to the rate of change of the wave energy.

The right-hand side of equation (\ref{1.17'''}) represents the work done on 
the original oscillations through non-linear couplings with a warp.
As mentioned before, there are four paths of the feedback to the 
original oscillations, i.e., two paths of $m$ $\rightarrow$ $m\pm 1$
$\rightarrow$ $m$ and two paths of $n$ $\rightarrow$ $n\pm 1$
$\rightarrow$ $n$.
We treat these four paths separately.
Equation (\ref{1.17'''}) shows that for the coupling of $m$ $\rightarrow$ 
$m+1$ $\rightarrow$ $m$, the rate of energy change of the oscillations 
is given by
\begin{equation}
    -4i{\omega_{\rm i}\over\omega_0}E = \Im
    \int\rho_0 \mbox{\boldmath $\xi$}^{(0)*}\mbox{\boldmath $C$}
        _{+,{\tilde n}}^{(2)}(\mbox{\boldmath $\zeta$}_{+,{\tilde n}}^{(1)},
                              \mbox{\boldmath $\xi$}^{{\rm w}*})dV,
\label{m+1}
\end{equation}
while  
\begin{equation}
    -4i{\omega_{\rm i}\over\omega_0}E = \Im
    \int\rho_0 \mbox{\boldmath $\xi$}^{(0)*}\mbox{\boldmath $C$}
        _{-,{\tilde n}}^{(2)}(\mbox{\boldmath $\zeta$}_{-,{\tilde n}}^{(1)},
                              \mbox{\boldmath $\xi$}^{\rm w})dV,
\label{m-1}
\end{equation}
when the path is $m$ $\rightarrow$ $m-1$ $\rightarrow$ $m$.
Here,  $\mbox{\boldmath $C$}_{+,{\tilde n}}^{(2)}$, for example,
represents the coupling between  
$\mbox{\boldmath $\xi$}_{+,{\tilde n}}^{(1)}$ and 
$\mbox{\boldmath $\xi$}^{{\rm w}*}$.
To make clear that this is the second stage of the coupling and that the path 
is $m$ $\rightarrow$ $m+1$ $\rightarrow$ $m$, the superscript $(2)$ and
the subscript $+$ have been attached to $\mbox{\boldmath $C$}$.

When integrations on the right-hand side of equations (\ref{m+1})
and (\ref{m-1}) are performed, we should first remember that among 
various terms of $\mbox{\boldmath $C$}^{(2)}$, we need only those
terms proportional to ${\cal H}_n(z/H)$ for $\mbox{\boldmath $C$}_\bot^{(2)}$ 
and to ${\cal H}_{n-1}(z/H)$ for $C_z^{(2)}$.
This is because $\mbox{\boldmath $\xi$}^{(0)}$ and $\xi_z^{(0)}$
are proportional to ${\cal H}_n(z/H)$ and ${\cal H}_{n-1}(z/H)$,
respectively, and Hermite polynomials of different orders are orthogonal
to each other.
Second, we should consider that $\mbox{\boldmath $\xi$}^{(1)}$'s have poles 
at the radii where $D=0$ is realized, and the imaginary parts of the
integrations (\ref{m+1}) and (\ref{m-1}) come from integration
around the poles.
Related to this, when we calculate $\mbox{\boldmath $C$}^{(2)}
(\mbox{\boldmath $\xi$}^{(1)},\mbox{\boldmath $\xi$}^{{\rm w}*})$,
the term $1/D$ in $\mbox{\boldmath $\xi$}^{(1)}$ can be moved outside 
of $\mbox{\boldmath $C$}^{(2)}$ as
$(1/D)\mbox{\boldmath $C$}^{(2)}(\mbox{\boldmath $\zeta$}^{(1)}, 
\mbox{\boldmath $\xi$}^{{\rm w}*})$.

Based on these considerations, we write the $\mbox{\boldmath $C$}^{(2)}$'s as
\begin{equation}
   \mbox{\boldmath $C$}_{\bot,+,{\tilde n}}^{(2)}={1\over D_{+,{\tilde n}}}
         {\cal H}_n(z/H){\rm exp}\{i(\omega t-m\varphi)\}
          \hat {\mbox {\boldmath $C$}}_{\bot,+,{\tilde n}}^{(2)}
         (\hat{\mbox{\boldmath $\zeta$}}_{+,{\tilde n}},
           \mbox{\boldmath $\eta$}^*),
\label{C2rtilde}
\end{equation}
\begin{equation}
   C_{z,+,{\tilde n}}^{(2)}={1\over D_{+,{\tilde n}}}
         {\cal H}_{n-1}(z/H){\rm exp}\{i(\omega t-m\varphi\}
          {\hat C}_{z,+,{\tilde n}}^{(2)}
         (\hat {\mbox{\boldmath $\zeta$}}_{+,{\tilde n}},
           \mbox{\boldmath $\eta$}^*).
\label{C2ztilde}
\end{equation}
Then, the right-hand side of equation (\ref{m+1}), for example, can be written
as 
\begin{equation}
  \biggr\{ (2\pi)^{3/2}rH\rho_{00}\biggr[
               n!\ \hat{\mbox{\boldmath $\xi$}}_\bot^{(0)*}
      \cdot \hat{\mbox{\boldmath $C$}}_{\bot,+,{\tilde n}}^{(2)}
     (\hat{\mbox{\boldmath $\zeta$}}_{+,{\tilde n}},
      \hat{\mbox{\boldmath $\eta$}}^*)
     +(n-1)!\ {\hat \xi}_z^{(0)*}
       \cdot\hat {C}_{z,+,{\tilde n}}^{(2)}(\hat{\mbox{\boldmath $\zeta$}}
                  _{+,{\tilde n}},
        \hat{\mbox{\boldmath $\eta$}}^{*})\biggr]
       \biggr\}_{\rm c}
      \Im\biggr(\int {dr\over D_{+,{\tilde n}}}\biggr),
\label{4.8}
\end{equation}
where the subscript c denotes the value at the resonance radius,
and the expressions for $\hat{\mbox{\boldmath $C$}}
_{\bot, +,{\tilde n}}^{(2)}$
and ${\hat C}_{z,+,{\tilde n}}^{(2)}$
are given in appendix 3.
The right-hand side of equation (\ref{m-1}) has the same form as
equation (\ref{4.8}), except that 
$\hat {C}_{\bot,+,{\tilde n}}^{(2)}(\hat{\mbox{\boldmath
$\zeta$}}_{+,{\tilde n}},\hat{\mbox{\boldmath $\eta$}^*})$
and 
$\hat{C}_{z,+,{\tilde n}}^{(2)}(\hat{\mbox{\boldmath
$\zeta$}}_{+,{\tilde n}},\hat{\mbox{\boldmath $\eta$}}^*)$ 
are replaced by 
$\hat {C}_{\bot,-,{\tilde n}}^{(2)}(\hat{\mbox{\boldmath
$\zeta$}}_{-,{\tilde n}},\hat{\mbox{\boldmath $\eta$}})$
and 
$\hat{C}_{z,-,{\tilde n}}^{(2)}(\hat{\mbox{\boldmath
$\zeta$}}_{-,{\tilde n}},\hat{\mbox {\boldmath $\eta$}})$.
Furthermore, $D_+$ in equation (\ref{4.8}) is replaced by
$D_-$.
Expressions for $\hat{\mbox{\boldmath $C$}}
_{\bot, -,{\tilde n}}^{(2)}$
and ${\hat C}_{z,-,{\tilde n}}^{(2)}$
are also given in appendix 3.

To perform integration around the poles,
we must be careful about the integration path,  
so that causality is satisfied (e.g., Kato 2003).
Let us consider the horizontal and vertical resonances separately.
For the horizontal resonances we have $[\omega-(m\pm 1)\Omega]^2-\kappa^2
\sim 0$ and around the resonance radius, $r_{\rm c}$, the first term of 
the Taylor expansion of $D_{\pm,{\tilde n}}$ gives
\begin{equation}
    D_{\pm,{\tilde n}}=2({\tilde n}\Omega_\bot^2-\kappa^2)
           G_{{\rm H},\pm}\biggr[{r-r_{{\rm c}}\over r}
         -i{\omega-(m\pm 1)\Omega\over G_{{\rm H},\pm}}
            \omega_{\rm i}\biggr],
\label{4.9}
\end{equation}
where
\begin{equation}
   G_{{\rm H},\pm}=\biggr\{(m\pm 1)\Omega[\omega-(m\pm 1)\Omega]
      {d{\rm ln}\Omega\over d{\rm ln}r}
      +\kappa^2{d{\rm ln}\kappa\over d{\rm ln}r}
     \biggr\}_{\rm c}.
\label{4.10}
\end{equation}
Considering these expressions and the requirement of causality, we have
\begin{equation}
   \int{1\over D_{\pm,{\tilde n}}}dr=
         -i\pi{r_{\rm c}\over 2({\tilde n}\Omega_\bot^2-\kappa^2)_{\rm c}
           \vert G_{{\rm H},\pm}\vert}
            {\rm sign}[\omega-(m\pm 1)\Omega]_{\rm c}.
\label{4.11}
\end{equation}
Here and hereafter, subscript c denotes the values at the resonance radius.

In the case of vertical resonances, $[\omega-(m\pm 1)\Omega]^2-{\tilde n}
\Omega_\bot^2\sim 0$ and around the resonance radius, $r_{\rm c}$, we have
\begin{equation}
    D_{\pm,{\tilde n}}=-2({\tilde n}\Omega_\bot^2-\kappa^2)
           G_{{\rm V},\pm,{\tilde n}}\biggr[{r-r_{{\rm c}}\over r}
         -i{\omega-(m\pm 1)\Omega\over G_{{\rm V},\pm,{\tilde n}}}
            \omega_{\rm i}\biggr],
\label{4.9'}
\end{equation}
where 
\begin{equation}
   G_{{\rm V},\pm,{\tilde n}}=\biggr\{(m\pm 1)\Omega[\omega-(m\pm 1)\Omega]
      {d{\rm ln}\Omega\over d{\rm ln}r}
      +2{\tilde n}\biggr(\Omega_\bot{d\Omega_\bot \over d{\rm ln}r}\biggr)
        \biggr\}_{\rm c}.
\label{4.10'}
\end{equation}
Similarly, we have
\begin{equation}
   \int{1\over D_{\pm,{\tilde n}}}dr=
         i\pi{r_{\rm c}\over 2({\tilde n}\Omega_\bot^2-\kappa^2)_{\rm c}
           \vert G_{{\rm V},\pm}\vert}
            {\rm sign}[\omega-(m\pm 1)\Omega]_{\rm c}.
\label{4.11'}
\end{equation}

Taking these considerations into account, we can write equation 
(\ref{m+1}) as
\begin{equation}
    -2\omega_{\rm i}=-{\pi\over 4(-\kappa^2+{\tilde n}\Omega^2)_{\rm c}}
                   \vert G_{H,+,{\tilde n}}\vert_{\rm c}
               {{\rm sign}[\omega-(m+1)\Omega]_{\rm c}
                \over (\omega-m\Omega)_{\rm c}}
               {{\tilde W}_{\rm c}\over {\tilde E}},
\label{4.12}
\end{equation}
where ${\tilde W}$ is related to the work done on oscillations, and is
\begin{equation}
  {\tilde W}=n!\ \hat{\xi}_{r,+,{\tilde n}}^{(0)}
                     \hat{C}_{r,+,{\tilde n}}^{(2)}
            +n!\ \hat{\xi}_{\varphi,+,{\tilde n}}^{(0)}
                     \hat{C}_{\varphi,+,{\tilde n}}^{(2)}
        +(n-1)!\ \hat{\xi}_{z,+,{\tilde n}}^{(0)}
                     \hat{C}_{z,+,{\tilde n}}^{(2)} 
\label{4.13}
\end{equation}
and
\begin{equation}
      {\tilde E}=\int {rH\rho_{00}\over r_{\rm c}H_{\rm c}\rho_{00}r_{\rm c}}
      {\omega-m\Omega\over \omega-m\Omega_{\rm c}}
        \biggr( n!\ \vert\hat{\xi}_r^{(0)}\vert^2+
        (n-1)!\ \vert\hat{\xi}_z^{(0)}\vert^2\biggr) {dr\over r_{\rm c}}.
\label{4.14}
\end{equation}
In the case of $m$ $\rightarrow$ $m-1$ $\rightarrow$ $m$, an expression for
the growth rate is similar to equation ({\ref{4.12}).
The differences are that the arguments of $\hat{\mbox{\boldmath $C$}}
_{-,{\tilde n}}^{(2)}$
are now $\hat{\mbox{\boldmath $\zeta$}}_{-,{\tilde n}}^{(1)}$ and
$\hat{\mbox{\boldmath $\eta$}}$.

Next, we should mention the case of vertical resonances.
Derivations of expressions for the growth rate are quite parallel with the 
case of horizontal resonances, and we have
\begin{equation}
    -2\omega_{\rm i}={\pi\over 4(-\kappa^2+{\tilde n}\Omega^2)_{\rm c}}
                   \vert G_{V,+,{\tilde n}}\vert_{\rm c}
               {{\rm sign}[\omega-(m+1)\Omega]_{\rm c}
                \over (\omega-m\Omega)_{\rm c}}
               {{\tilde W}_{\rm c}\over {\tilde E}}.
\label{4.15}
\end{equation}
We should especially notice that the sign is different from equation 
(\ref{4.12}).

\section{Growth Rate of Resonant Oscillations}

In this section, the growth (or damping) rates of resonant oscillations
are calculated using equations (\ref{4.12}) and (\ref{4.15}).
As discussed in section 5, there are three kinds of resonances:
i) horizontal and ii) vertical resonances of g-mode 
oscillations and iii) horizontal resonances of p-mode oscillations.
These resonances are discussed here separately.
It is noted that the oscillation modes of $n=0$ are p-modes alone.
These p-modes are similar in dynamical behaviors with g-mode
oscillations of $n\geq 1$.
Hence, the p-modes of $n=0$ are treated hereafter together with g-mode 
oscillations of $n\geq 1$.
In other words, when we mention p-mode oscillations, they are 
p-mode oscillations of $n\geq 1$.
Two basic approximations are introduced here.
The first one is that $c_{\rm s}^2k^2$ is much smaller than the square of
the rotational frequency, $\Omega^2$, which means that we adopt
$(kH)^2\ll 1$.
This is a natural approximation in treating disk oscillations in 
geometrically thin disks.
Introduction of this approximation implies that we are interested in
resonances that occur near to both boundaries of the propagation and
resonance regions.
That is, we consider resonances where
$[(\omega-m\Omega)^2-\kappa^2][(\omega-m\Omega)^2-n\Omega_\bot^2]\sim 0$
and $\{[\omega-(m\pm 1)\Omega]^2-\kappa^2\}\{[\omega-(m\pm 1)\Omega]^2
-{\tilde n}\Omega_\bot^2\}\sim 0$ are simultaneously realized.
The second approximation we adopt here is a local approximation of
$(kr)^2\gg 1$.
This approximation is relevant, except for just the 
boundaries of the resonance and progation regions, since we are treating 
geometrically thin disks.

\subsection{G-Mode Oscillations}

First, we summarize the relations among $\hat{\xi}_r^{(0)}$,
$\hat{\xi}_\varphi^{(0)}$, and $\hat{\xi}_z^{(0)}$
in g-mode oscillations.
In g-mode oscillations, $(\omega-m\Omega)^2-\kappa^2$ is close to
zero, but can be approximated as 
$(kH)^2\Omega_\bot^2[\kappa^2/(\kappa^2-n\Omega_\bot^2)]$ by using the
dispersion relation (\ref{disp}).
Hence, from equation (\ref{1.26c}) we have a relation between 
$\hat{\xi}_r^{(0)}$ and $\hat{\xi}_z^{(0)}$ as
\begin{equation}
      \hat{\xi}_z^{(0)}=-i(kH){n\Omega_\bot^2\over \kappa^2-n\Omega_\bot^2}
                         \hat{\xi}_r^{(0)}.
\label{7.1}
\end{equation}
A relation between $\hat{\xi}_\varphi^{(0)}$ and $\hat{\xi}_r^{(0)}$ is,
from equation ({\ref{1.26b}),
\begin{equation}
    \hat{\xi}_\varphi^{(0)}=i{2\Omega\over \omega-m\Omega}
     \hat{\xi}_r^{(0)}.
\label{7.2}
\end{equation}
These relations show that in g-mode oscillations, $\hat{\xi}_z^{(0)}$
is smaller than $\hat{\xi}_r^{(0)}$ and $\hat{\xi}_\varphi^{(0)}$ by
a factor of $(kH)$.

Since in the paths of $n$ $\rightarrow$ $n+1$ $\rightarrow$ $n$ 
(i.e., ${\tilde n}=n+1$) and $n$ $\rightarrow$ $n-1$ $\rightarrow$ $n$ 
(i.e., ${\tilde n}=n-1$), expressions for $\hat{\mbox{\boldmath $C$}}^{(1)}$'s
are different, we treat these two cases separately.
In the case of the upper path (${\tilde n}=n+1$), negelcting small-order 
quantities, we have from appendix 2
\begin{equation}
    \hat{C}_{r,+,n+1}^{(1)}
      =i(kH)\Omega_\bot^2\hat{\xi}_r^{(0)}
     \biggr({d\eta\over dr}+{2\Omega\over \omega-m\Omega}{\eta\over r}\biggr),
\label{7.3}
\end{equation}
\begin{equation}
    \hat{C}_{z,+,n+1}^{(1)}
      =\Omega_\bot^2\hat{\xi}_r^{(0)}
       \biggr[ -{d{\rm ln}\Omega_\bot^2\over dr}\eta+
           n\biggr({d\eta\over dr}+{2\Omega\over \omega-m\Omega}
                               {\eta\over r}\biggr)\biggr].
\label{7.4}
\end{equation}
Here, we do not present $\hat{C}_{\varphi,+,n+1}^{(1)}$,
since it is smaller than $\hat{C}_{r,+,n+1}^{(1)}$
and unnecessary hereafter.
Although $\hat{C}_{r,+,n+1}^{(1)}$ is smaller than
$\hat{C}_{z,+,n+1}^{(1)}$ by a factor of $kH$, it contributes 
to the final results.

In the case of the lower path, on the other hand, using expressions in
appendix 2 and approximations of $kH\ll 1$ and $kr\gg 1$, we have
\begin{equation}
   \hat{C}_{r,+,n-1}^{(1)}=
    \Omega_\bot^2
      \biggr\{ 
           -in(kH)\hat{\xi}_r^{(0)}{d\eta\over dr}
             +\hat{\xi}_z^{(0)}
          \biggr(-{d{\rm ln}\Omega_\bot^2\over dr}\eta+
           (n-1){d\eta\over dr}\biggr) 
       \biggr\},
\label{7.5}
\end{equation}
\begin{equation}
   \hat{C}_{\varphi,+,n-1}^{(1)}=
    -\Omega_\bot^2{1\over r}\biggr[n(kH)\hat{\xi}_r^{(0)}
             +i(n-1)\hat{\xi}_z^{(0)}\biggr]\eta.
\label{7.6}
\end{equation}
The term $\hat{C}_{z,+,n-1}^{(1)}$ is not presented
here since it is smaller than $\hat{C}_{r,+,n-1}^{(1)}$
and $\hat{C}_{\varphi,+,n-1}^{(1)}$ by a factor of
$H/r$, and is unnecessary hereafter.

\subsubsection{Horizontal resonances of g-mode oscillations}

To proceed further, we separately consider two coupling paths of 
$n$ $\rightarrow$ 
$n+1$ $\rightarrow$ $n$ and $n$ $\rightarrow$ $n-1$ $\rightarrow$ $n$.

\noindent
(a) Coupling through the upper path (${\tilde n}=n+1$)

In this case, we can express $\hat{\mbox{\boldmath
$\zeta$}}_{+,n+1}^{(1)}$ in terms of 
$\hat{\mbox{\boldmath $\xi$}}_{+,n+1}^{(1)}$ by using equations
(\ref{2.10})--(\ref{2.12}) under the help of equations (\ref{7.3}) and
(\ref{7.4}).
The results are
\begin{equation}
    \hat{\zeta}_{r,+,n+1}^{(1)}
      =i(kH)\Omega_\bot^4\hat{\xi}_r^{(0)}
       \biggr\{{d{\rm ln}\Omega_\bot^2\over dr}\eta
           +{-\kappa^2+\Omega_\bot^2\over \Omega_\bot^2}
              \biggr({d\eta\over dr}+{2\Omega\over \omega-m\Omega}
                {\eta\over r}\biggr)\biggr\},
\label{7.7}
\end{equation}
\begin{equation}
    \hat{\zeta}_{\varphi,+,n+1}^{(1)}
      =i{2\Omega\over \omega-(m+1)\Omega}\hat{\zeta}_{r,+,n+1}^{(1)},
\label{7.8}
\end{equation}
\begin{equation}
    \hat{\zeta}_{z,+,n+1}^{(1)}
      =i{\tilde n}(kH){\Omega_\bot^2\over -\kappa^2+{\tilde n}\Omega_\bot^2}
          \hat{\zeta}_{r,+,n+1}^{(1)}.
\label{7.9}
\end{equation}
This shows that $\hat{\zeta}_{z,+,n+1}^{(1)}$ is smaller that
$\hat{\mbox {\boldmath $\zeta$}}_{\bot,+,n+1}^{(1)}$ by a factor  of
$kH$, but it is necessary in deriving ${\tilde W}$.

The next step is to calculate $\hat{\mbox{\boldmath $C$}}_{+,n+1}^{(2)}$
by using equations (\ref{7.7})--(\ref{7.9}).
By consulting the expressions for $\hat{\mbox{\boldmath $C$}}(
\hat{\mbox{\boldmath $\xi$}},
\hat{\mbox{\boldmath $\xi$}})$
given in appendix 2, we can derive $\hat{\mbox{\boldmath $C$}}
(\hat{\mbox{\boldmath $\xi$}}^{(1)},
\hat{\mbox{\boldmath $\eta$}}^*)$.
Considering the order of each term, we have
\begin{equation}
    \hat{C}_{r,+,n+1}^{(2)}
      =-i{\tilde n}(kH)\Omega_\bot^2\hat{\zeta}_{r,+,n+1}^{(1)}
            {d\eta^*\over dr}+
             \Omega_\bot^2\hat{\zeta}_{z,+,n+1}^{(1)}
             \biggr(
                -{d{\rm ln}\Omega_\bot^2\over dr}\eta^*
                +n{d\eta^*\over dr}\biggr),
\label{7.10}
\end{equation}
\begin{equation}
    \hat{C}_{\varphi,+,n+1}^{(2)}
      ={\Omega_\bot^2 \over r}
        \biggr
      [{\tilde n}(kH)\hat{\zeta}_{r,+,n+1}^{(1)}+in\hat{\zeta}_{z,+,n+1}^{(1)}
          \biggr]\eta^*.
\label{7.11}
\end{equation}
Here, $\hat{C}_{z,+,n+1}^{(2)}$ is omitted, since it
does not contribute to ${\tilde W}$.
Using the above expressions, we finally have, after lengthy
manipulations, a simple expression:
\begin{equation}
   {\tilde W}=n!\biggr(\hat{\xi}_r^{(0)}\hat{C}_{r,+,n+1}^{(2)}
          +\hat{\xi}_\varphi^{(0)}\hat{C}_{\varphi,+,n+1}^{(2)}\biggr)
        ={{\tilde n}!\over -\kappa^2+{\tilde n}\Omega_\bot^2}
           \vert \hat{\zeta}_{r,+,n+1}^{(1)}\vert^2.
\label{7.12}
\end{equation}

\noindent
(b) Coupling through the lower path (${\tilde n}=n-1$)

In this case, $\hat{C}_{r,+,n-1}^{(1)}$ and 
$\hat{C}_{\varphi,+,n-1}^{(1)}$ are given by equations (\ref{7.5}) and
(\ref{7.6}).
Hence, in the lowest order of approximations we have, from equations 
(\ref{2.10}) and (\ref{2.11}), 
\begin{eqnarray} 
   \hat{\zeta}_{r,+,n-1}^{(1)}&=&-in(kH)\Omega_\bot^4
            {-\kappa^2+{\tilde n}\Omega_\bot^2  \over
             -\kappa^2+n\Omega_\bot^2}\hat{\xi}_r^{(0)}  \nonumber \\
      & & \times\biggr\{{d{\rm ln}\Omega_\bot^2\over dr}\eta
          +{-\kappa^2+\Omega_\bot^2\over \Omega_\bot^2}
           \biggr[{d\eta\over dr}-{2\Omega\over \omega-(m+1)\Omega}
           {\eta\over r}\biggr]\biggr\},
\label{7.14}
\end{eqnarray}
\begin{equation}
    \hat{\zeta}_{\varphi,+,n-1}^{(1)}=i{2\Omega\over \omega-(m+1)\Omega}
                         \hat{\zeta}_{r,+,n-1}^{(1)}.
\label{7.13}
\end{equation}
In deriving ${\tilde W}$, we do not need $\hat{\zeta}_{z,+,n-1}^{(1)}$.

Using the above expressions, we can obtain 
$\hat{\mbox{\boldmath $C$}}_{+,n-1}^{(2)}$.
The results are summarized as
\begin{equation}
    \hat{C}_{r,+,n-1}^{(2)}=i(kH)\Omega_\bot^2\hat{\zeta}_{r,+,n-1}^{(1)}
         \biggr[{d\eta^*\over dr}
          -{2\Omega\over \omega-(m+1)\Omega}{\eta^*\over r}\biggr],
\label{7.15}
\end{equation}
\begin{equation}
    \hat{C}_{z,+,n-1}^{(2)}=\Omega_\bot^2\hat{\zeta}_{r,+,n-1}^{(1)}
        \biggr\{-{d{\rm ln}\Omega_\bot^2\over dr}\eta^*
        +(n-1)\biggr[{d\eta^*\over dr}
          -{2\Omega\over \omega-(m+1)\Omega}{\eta^*\over r}\biggr]\biggr\}.
\label{7.16}
\end{equation}
The $\varphi$-component, $\hat{C}_{\varphi,+,n-1}^{(2)}$,
is not presented here, since it is smaller than
$\hat{C}_{r,+,n-1}^{(2)}$
and does not contribute to ${\tilde W}$.
Taking into account these results, we finally have, after some
manipulations,
\begin{equation}
   {\tilde W}=n!\ \xi_{r,+,n-1}^{(0)*}\hat{C}_{r,+,n-1}^{(2)}
        +(n-1)!\ \xi_{z,+,n-1}^{(0)*}\hat{C}_{z,+,n-1}^{(2)}
      ={{\tilde n}!\over -\kappa^2+{\tilde n}\Omega_\bot^2}
        \vert\zeta_{r,+,n-1}^{(1)}\vert ^2.
\label{7.18}
\end{equation}

\subsubsection{Vertical resonances of g-mode oscillations}

\noindent
(a) Coupling through the upper path (${\tilde n}=n+1$)

Expressions for $\hat{C}_{r,+,n+1}^{(1)}$ and $\hat{C}_{z,+,n+1}^{(1)}$
are the same as those for the horizontal resonances given by equations 
(\ref{7.3}) and (\ref{7.4}).
In the present case of vertical resonances, however, expressions for the 
lowest order of approximations of 
$\hat{\mbox{\boldmath $\zeta$}}_{+,n+1}^{(1)}$ 
are different from those in the horizontal resonances.
In the expression for $\hat{\zeta}_{r,+,{\tilde n}}^{(1)}$ 
[equation (\ref{2.10})], the terms coming from $\hat{C}_{r,+,n+1}^{(1)}$
and $\hat{C}_{\varphi,+,n+1}^{(1)}$ are smaller than the term of 
$\hat{C}_{z,+,n+1}^{(1)}$, since $[\omega-(m+1)\Omega]^2-{\tilde
n}\Omega_\bot^2$ is now small by the resonance being vertical.
The results of calculations are
\begin{eqnarray}
  \hat{\zeta}_{r,+,n+1}^{(1)}&=&-i(kH)\Omega_\bot^2\hat{C}_{z,+,n+1}^{(1)}
                              \nonumber \\
      &=& -i(kH)\Omega_\bot^4\hat{\xi}_r^{(0)}
        \biggr[ -{d{\rm ln}\Omega_\bot^2\over dr}\eta+
               n\biggr({d\eta\over dr}+{2\Omega\over \omega-m\Omega}
                    {\eta\over r}\biggr)\biggr],
\label{7.21}
\end{eqnarray}
\begin{equation}
  \hat{\zeta}_{z,+,n+1}^{(1)}=i(kH)^{-1}
         {\kappa^2-{\tilde n}\Omega_\bot^2 \over \Omega_\bot^2}
            \hat{\zeta}_{r,+,n+1}^{(1)},
\label{7.23}
\end{equation}
where the expression for $\hat{\zeta}_{\varphi,+,n+1}^{(1)}$ is not 
presented, since it is the same as equation (\ref{7.8}).
These equations show that, different from the case of horizontal 
resonances, the intermediate
oscillations have the largest amplitude in the vertical direction.

By using $\hat{\mbox{\boldmath $\zeta$}}_{+,n+1}^{(1)}$ given above, we
can calculate $\hat{\mbox{\boldmath $C$}}_{+,n+1}^{(2)}$.
The results show that the main terms of $\hat{\mbox{\boldmath $C$}}
_{+,n+1}^{(2)}$ are
\begin{equation}
    \hat{C}_{r,+,n+1}^{(2)}=\Omega_\bot^2\hat{\zeta}_{z,+,n+1}^{(1)}
       \biggr(-{d{\rm ln}\Omega_\bot^2 \over dr}\eta^*+n{d\eta^*\over dr}
           \biggr),
\label{7.24}
\end{equation}
\begin{equation}
    \hat{C}_{\varphi,+,n+1}^{(2)}=in\Omega_\bot^2\hat{\zeta}_{z,+,n+1}^{(1)}
       {\eta^*\over r}.
\label{7.25}
\end{equation}
The $z$-component, $\hat{C}_{z,+,n+1}^{(2)}$, is found to be unnecessary 
here.
From these results, we finally have
\begin{equation}
     \tilde {W}=n!\biggr[\hat{\xi}_r^{(0)}\hat{C}_{r,+,n+1}^{(2)}
          +\hat{\xi}_\varphi^{(0)}\hat{C}_{\varphi,+,n+1}^{(2)}\biggr]
        =-{({\tilde n}-2)!\over -\kappa^2+{\tilde n}\Omega_\bot^2}
              \vert\hat{\zeta}_{z,+,n+1}^{(1)}\vert^2.
\label{7.26}
\end{equation}

\noindent
(b) Coupling through the lower path (${\tilde n}=n-1$)

The expressions for $\hat{C}_{r,+,n-1}^{(1)}$ and 
$\hat{C}_{\varphi,+,n-1}^{(1)}$ are the same as those given by equations 
(\ref{7.5}) and (\ref{7.6}), respectively, and
$\hat{C}_{z,+,n-1}^{(1)}$ is unnecessary, as in the case of (b) of
subsubsection 7.1.1.
Hence, all of the procedures go on in parallel, like in subsubsection 7.1.1.
In subsubsection 7.1.1. in deriving  $\hat{\zeta}_{r,+,n+1}^{(1)}$ from
equation (\ref{2.10}), $[\omega-(m+1)\Omega]^2-{\tilde n}\Omega_\bot^2$ was
replaced by $\kappa^2-{\tilde n}\Omega_\bot^2$, since the horizontal
resonances are considered there.
In the present case of vertical resonances, it can be approximated 
by ${\tilde n}(kH)^2\Omega_\bot^4/({\tilde n}\Omega_\bot-\kappa^2)$.
Hence, in the present case we have
\begin{eqnarray}
     \hat{\zeta}_{r,+,n-1}^{(1)}
      &=&-(kH)^2{{\tilde n}\Omega_\bot^4\over {\tilde n}\Omega_\bot^2-\kappa^2}
          \biggr[\hat{C}_{r,+,n-1}^{(1)}-i
       {2\Omega\over \omega-(m+1)\Omega}\hat{C}_{\varphi,+,n-1}^{(1)}\biggr]
                     \nonumber \\
      &=& i(kH)^3{n{\tilde n}\Omega_\bot^8\over 
                (-\kappa^2+{\tilde n}\Omega_\bot^2)
                (-\kappa^2+n\Omega_\bot^2)}\hat{\xi}_r^{(0)}
                      \nonumber \\
      &\times&
        \biggr\{{d{\rm ln}\Omega_\bot^2\over dr}\eta
          +{-\kappa^2+\Omega_\bot^2 \over \Omega_\bot^2}
         \biggr[{d\eta\over dr}-{2\Omega\over \omega-(m+1)\Omega}
       {\eta\over r}\biggr]\biggr\}, 
\label{7.27}
\end{eqnarray}
\begin{equation}
   \hat{\zeta}_{\varphi,+,n-1}^{(1)}=i{2\Omega\over \omega-(m+1)\Omega}
                    \hat{\zeta}_{r,+,n-1}^{(1)},
\label{7.28}
\end{equation}
\begin{equation}
   \hat{\zeta}_{z,+,n-1}^{(1)}=i(kH)^{-1}
   {n\Omega_\bot^2-\kappa^2\over \Omega_\bot^2}\hat{\zeta}_{r,+,n-1}^{(1)}.
\label{7.28'}
\end{equation}
Expressions for $\hat{C}_{r,+,n-1}^{(2)}$ and $\hat{C}_{z,+,n-1}^{(2)}$
are the same as those given by equations (\ref{7.15}) and (\ref{7.16}).
Hence, we finally have
\begin{equation}
    {\tilde W}=
    n!\ \hat{\xi}_{r,+,n-1}^{(0)}\hat{C}_{r,+,n-1}^{(2)}+
    (n-1)!\ \hat{\xi}_{z,+,n-1}^{(0)}\hat{C}_{z,+,n-1}^{(2)}
      =-{({\tilde n}-2)!\over -\kappa^2+{\tilde n}\Omega_\bot^2}
       \vert \hat{\zeta}_{z,+,n-1}^{(1)}\vert ^2.
\label{7.31}
\end{equation}

\subsection{P-Mode Oscillations}

In the case of p-mode oscillations,
$-(\omega-m\Omega)+n\Omega_\bot^2\sim 0$, but the dispersion relation
shows that it can be approximated as 
$-n(kH)^2\Omega_\bot^4/(n\Omega_\bot^2-\kappa^2)$.
Hence, from equation (\ref{1.26c}) we have
 a relation between $\hat{\xi}_z^{(0)}$ and $\hat{\xi}_r^{(0)}$, which
is
\begin{equation}
   \hat{\xi}_r^{(0)}=i(kH){\Omega_\bot^2\over n\Omega_\bot^2-\kappa^2}
              \xi_z^{(0)}.
\label{7.41}
\end{equation}
Furthermore, equation (\ref{1.26b}) gives
\begin{equation}
     \hat{\xi}_\varphi^{(0)}=i{2\Omega\over \omega-m\Omega}\xi_r^{(0)}.
\label{7.42}
\end{equation}
These relations show that in the p-mode oscillations $\hat{\xi}_z^{(0)}$ is
major and that $\hat{\xi}_r^{(0)}$ and $\hat{\xi}_\varphi^{(0)}$ 
are smaller than 
$\hat{\xi}_z^{(0)}$ by a factor of $kH$, as expected.

As mentioned before, we have only horizontal resonances, i.e., the
vertical resonances are absent in the case of p-mode oscillations.

\subsubsection{Horizontal resonances of p-mode oscillations}

As in the case of g-mode oscillations, two coupling paths are treated
separately.

\noindent
(a) Coupling through the upper path (${\tilde n}=n+1$)

Expressions for $\hat{C}_{r,+,n+1}^{(1)}$ and $\hat{C}_{z,+,n+1}^{(1)}$
are the same as equations (\ref{7.3}) and (\ref{7.4}), respectively.
($\hat{C}_{\varphi,+,n+1}^{(1)}$ is unnecessary, since it is small.)
The expressions for $\hat{\zeta}_{r,+,n+1}^{(1)}$ and 
$\hat{\zeta}_{z,+,n+1}^{(1)}$ are also the same as those given by
equations (\ref{7.7}) and (\ref{7.9}), respectively.
The coupling terms, $\hat{C}_{r,+,n+1}^{(2)}$ and $\hat{C}_
{\varphi,+,n+1}^{(2)}$ are thus the same as those given by equations 
(\ref{7.10}) and (\ref{7.11}).
The term $\hat{C}_{z,+,n+1}^{(2)}$ is unnecessary, since it is small and
does not contribute to the final results.

Using the above results, we finally have
\begin{equation}
    {\tilde W}=n!\biggr[\hat{\xi}_r^{(0)}\hat{C}_{r,+,n+1}^{(2)}
       +\hat{\xi}_\varphi^{(0)}\hat{C}_{\varphi,+,n+1}^{(2)}\biggr]
     ={{\tilde n}!\over -\kappa^2+{\tilde n}\Omega_\bot^2}
      \vert \hat{\zeta}_{r,+,n+1}^{(1)}\vert^2.
\label{7.47}
\end{equation}
This expression for ${\tilde W}$ is the same as equation 
(\ref{7.12}), but its concrete figure is different from that of equation 
(\ref{7.12}), since the expressions for $\hat{\xi}_r^{(0)}$ and 
$\hat{\xi}_z^{(0)}$ are different from those 
of the g-mode oscillations (see table 1).

\noindent
(b) Coupling through the lower path (${\tilde n}=n-1$)

The coupling terms in the first stage are found to be
\begin{equation}
  \hat{C}_{r,+,n-1}^{(1)}=\Omega_\bot^2\hat{\xi}_{z}^{(0)}
         \biggr[-{d{\rm ln}\Omega_\bot^2\over dr}\eta
         +{\tilde n}{d\eta\over dr}\biggr],
\label{7.48}
\end{equation}
\begin{equation}
  \hat{C}_{\varphi,+,n-1}^{(1)}=-i{\tilde n}\Omega_\bot^2
           \hat{\xi}_{z}^{(0)}
             {\eta\over r}.
\label{7.49}
\end{equation}
These expressions for $\hat{C}_{r,+,n-1}^{(1)}$ and 
$\hat{C}_{\varphi,+,n-1}^{(1)}$ are different from equations 
(\ref{7.5}) and (\ref{7.6}), respectively, since the terms of 
$\hat{\xi}_r^{(0)}$ can be neglected in the present case.
 
Using the above expressions, we have
\begin{equation}
     \hat{\zeta}_{r,+,n-1}^{(1)}
        =(-\kappa^2+{\tilde n}\Omega_\bot^2)\Omega_\bot^2\hat{\xi}_z^{(0)}
         \biggr\{-{d{\rm ln}\Omega_\bot^2\over dr}\eta
         +{\tilde n}\biggr[{d\eta\over dr}
        -{2\Omega\over \omega-(m+1)\Omega}{\eta\over r}\biggr]\biggr\}.
\label{7.49}
\end{equation}
Expressions for $\hat{\zeta}_{\varphi,+,n-1}^{(1)}$ and 
$\hat{\zeta}_{z,+,n-1}^{(1)}$ are not shown here, since they are the same as 
equations (\ref{7.8}) and (\ref{7.9}), respectively, except for the
subscript $n+1$ being replaced by $n-1$.
Consequently, as the final results concerning ${\tilde W}$, we have
\begin{equation}
      {\tilde W}=(n-1)!\ \hat{\xi}_z^{(0)}\hat{C}_{z,+,n-1}^{(2)}
       ={{\tilde n}! \over -\kappa^2+{\tilde n}\Omega_\bot^2}
         \vert \hat{\zeta}_{r,+,n-1}^{(1)}\vert^2.
\label{7.52}
\end{equation}

So far, we have restricted our attention to the coupling cases of
$m$ $\rightarrow$ $m+1$ $\rightarrow$ $m$.
Even in the case of $m$ $\rightarrow$ $m-1$ $\rightarrow$ $m$,
however, the derivation of the ${\tilde W}$'s is similar.
The results show that the final expressions for ${\tilde W}$'s are the
same, except that the subscript + in $\hat{\zeta}_{r,+,n+1}^{(1)}$, 
$\hat{\zeta}_{r,+,n-1}^{(1)}$,
$\hat{\zeta}_{z,+,n+1}^{(1)}$, or $\hat{\zeta}_{z,+,n-1}^{(1)}$
in equations (\ref{7.12}), (\ref{7.18}), (\ref{7.26}), (\ref{7.31}),
(\ref{7.47}), or (\ref{7.52}) is replaced by the subscript $-$.
The difference between $\hat{\zeta}_{r,+,n+1}^{(1)}$ and 
$\hat{\zeta}_{r,-,n+1}^{(1)}$, for example, is  
that $m+1$ in the expressions for $\hat{\zeta}_{r,+,n+1}^{(1)}$ is
changed to $m-1$ in $\hat{\zeta}_{r,-,n+1}$.

Next, we summarize the expressions of growth rate.
As mentioned in section 5, the sign of $[\omega-(m\pm 1)\Omega]
/(\omega-m\Omega)$
is negative in the horizontal resonances of g-mode oscillations.
Hence, in this case we have, from equation (\ref{4.12}) and expressions for
$\hat{W}$'s given above,
\begin{equation}
   -\omega_{\rm i} ={\pi/8\over (-\kappa^2+{\tilde n}\Omega^2)^2}
      \ \biggr\vert {G_{H,\pm,{\tilde n}}\over \omega-m\Omega}\biggr\vert_{\rm c}
      \ {{\tilde n}!\ \vert\hat{\zeta}_{r,\pm,{\tilde n}}^{(1)}\vert_{\rm c}^2
        \over \hat{E}},
\label{7.1.1}
\end{equation}
which is positive, and thus the resonant oscillations are amplified
($-\omega_{\rm i}>0$).
In the cases of horizontal resonances of p-mode oscillations (excluding
$n=0$), on the other hand, the sign of 
$[\omega-(m\pm 1)\Omega]/(\omega-m\Omega)$ at the
resonant point is positive, and we have
\begin{equation}
   -\omega_{\rm i} =-{\pi/8\over (-\kappa^2+{\tilde n}\Omega^2)^2}
      \ \biggr\vert {G_{H,\pm,{\tilde n}}\over \omega-m\Omega}\biggr\vert_{\rm c}
      \ {{\tilde n}!\ \vert\hat{\zeta}_{r,\pm,{\tilde n}}^{(1)}\vert^2_{\rm c}
         \over \hat{E}},
\label{7.1.2}
\end{equation}
which is negative and the resonances lead to damping of the oscillations.

In the case of vertical resonances of g-mode oscillations,
the situations are a little different.
The sign of $[\omega-(m\pm 1)\Omega]/(\omega-m\Omega)$ is positive
in this case.
In the case of vertical resonances, the sign of $\int(1/D)dr$ is
opposite to that in the case of horizontal resonances [see equations
(\ref{4.11}) and (\ref{4.11'})].
The signs of the $\hat{W}$'s given above are also opposite to those in the
case of horizontal resonances.
Considering them, we have
\begin{equation}
   -\omega_{\rm i} =-{\pi/8\over (-\kappa^2+{\tilde n}\Omega^2)^2}
      \ \biggr\vert {G_{V,\pm,{\tilde n}}\over \omega-m\Omega}\biggr\vert_{\rm c}
      \ {({\tilde n}-2)!\ \vert\hat{\zeta}_{z,\pm,{\tilde n}}^{(1)}\vert^2_{\rm c}
        \over \hat{E}},
\label{7.1.3}
\end{equation}
and the oscillations are damped.

\section {Discussion}

In this paper the resonant excitations of disk oscillations on 
relativistic disks deformed by a warp have been examined, considering 
non-linear couplings between the oscillations and the warp.
The non-linear couplings that we consider are schematically shown
in figure 1.
The resonant processes involved in the couplings bring about amplification, or
damping, of the original disk oscillations.
First, a general stability criterion of the resonances is derived by using the
Lagrangian formulation developed by Lynden-Bell and Ostriker (1974).
Then, the criterion is applied to studying the resonant amplification.

There are three types of resonances: i) horizontal and ii) vertical
resonances of g-mode oscillations and iii) vertical resonances of
p-mode oscillations.
For each case, four coupling paths between the oscillations and 
the warp are possible, i.e., $m$ $\rightarrow$ $m\pm 1$ $\rightarrow$
$m$ and $n$ $\rightarrow$ $n\pm 1$ $\rightarrow$ $n$.
Detailed examinations where the resonances occur and how much the 
frequencies of the resonant oscillations are were made in Paper I.
In this paper we thus concentrate on whether these resonances amplify 
or dampen the oscillations.
The results concerning their growth rates are summarized in equations
({\ref{7.1.1})--(\ref{7.1.3}).
It should be emphasized that the presence of resonances is   
completely an effect of general relativity.
This is because a difference in the spatial distributions of $\Omega(r)$
and $\kappa(r)$ is essential for the resonances, and it results from the
disk being general relativistic.

Although the detailed interaction processes are different in each 
type of resonances and in each coupling path, the final expressions
for ${\tilde W}$ have similar compact forms for all cases [see
equations (\ref{7.12}), (\ref{7.18}), (\ref{7.26}), (\ref{7.31}), 
(\ref{7.47}), and (\ref{7.52})], which suggests the correctness of our 
manipulations.
[The results of Kato (2003) are not always correct, since there are some
errors in the calculations.] 

In the case of horizontal resonances, ${\tilde W}$ can be compactly 
expressed in terms of $\hat{\zeta}_r^{(1)}$, while it is expressed
in terms of $\hat{\zeta}_z^{(1)}$ in the case of vertical resonances,
as expected.
Although the expressions for ${\tilde W}$ are similar in all cases, concrete
figures of the growth rates are different in each case, since the expressions
for ${\hat\zeta}_r^{(1)}$ (or $\hat{\zeta}_z^{(1)}$) are different
in each case.
Let us now estimate the order of the growth (or damping) rate.
The amplitudes of intermediate oscillations, $\hat{\mbox{\boldmath $\zeta$}}
^{(1)}$, are given in section 7 for various types of resonances.
Hence, using them we can estimate the order of the growth rates, which
are summarized in table 1.
In table 1, $\alpha$ denotes $\vert\eta\vert_{\rm c}^2/r_{\rm c}L$,
where $L$ is the radial size of the wave packet of the oscillations,
and $\hat{E}$ has been estimated by 
$\hat{E}\sim \vert\hat{\xi}_r^{(0)}\vert^2L/r$ for g-mode oscillations 
and by $\hat{E}\sim \vert\hat{\xi}_z^{(0)}\vert^2L/r$ for p-mode oscillations.
The radii where the resonances for the oscillations of $k\sim 0$ occur 
are also given in table 1 (see also Paper I).
It is noted that in table 1 p-mode oscillations of $n=0$ are included 
in g-mode oscillations.

\begin{longtable}{llll}
  \caption{Growth rates and resonant radii.}\label{tab:LTsample}
  \hline\hline
    Type of resonance & Path & Growth rate & Radii \\
  \hline
\endfirsthead
\endhead
\hline
\endlastfoot
  g-mode, horizontal & ${\tilde n}=n\pm 1$ 
                     & $\ \  \alpha(kH)^2\Omega$
                     & $\sim 4.0r_{\rm g}$ \\
  g-mode, vertical   & ${\tilde n}=n+1$    & $-\alpha\Omega$ (damping)
                     & $\sim 3r_{\rm g}$, $3.62r_{\rm g}$, $6.46r_{\rm g}$ \\
                     & ${\tilde n}=n-1$    & $-\alpha(kH)^4\Omega$
                     & $\sim 3r_{\rm g}$, $3.62r_{\rm g}$, $6.46r_{\rm g}$ \\
  p-mode, horizontal & ${\tilde n}=n+1$    & $-\alpha(kH)^4\Omega$ (damping)
                     & $\sim 3r_{\rm g}$, $3.62r_{\rm g}$, $6.46r_{\rm g}$ \\
  ($n\not =0$)       & ${\tilde n}=n-1$,   & $-\alpha\Omega$ (damping)
                     & $\sim 3r_{\rm g}$, $3.62r_{\rm g}$, $6.46r_{\rm g}$ \\

\end{longtable}
\footnotetext{$\alpha$ in table 1 denotes $\vert\eta\vert_{\rm c}^2
                     /r_{\rm c}L$.}

The radii given in table 1 are those where both curves
specifying the propagation region and the resonance region cross on the
$\omega$-$r$ plane.
In other words, they are radii where long-wavelength oscillations have
resonances.
We think that long-wavelength oscillations are of interest, since they
are most observable (little phase mixing and little group velocity in
the radial direction).
The growth (or damping) rate of the oscillations, however, tends to 
zero in the limit of long wavelength of $kH=0$.
Hence, the oscillations that are of interest are those slightly apart 
from the radii given in the table.

Table 1 shows that the vertical resonances of the g-mode oscillations 
with ${\tilde n}=n+1$ and the horizontal resonances of the p-mode 
oscillations with ${\tilde n}=n-1$ are strongly damped 
[see the absence of a small factor of $(kH)$].
The growing modes are horizontal resonances of the g-mode oscillations alone,
including the horizontal resonances of the p-mode oscillations of $n=0$.
The growth rates of these oscillations are on the order of
$\alpha\Omega_{\rm c}$ times $(kH)^2$.
This suggests that the quasi-periodic oscillations caused by our present 
mechanism are observable when the disks are moderately thick.

Let us hereafter focus only on the growing oscillations, which are all
amplified around $4.0r_{\rm g}$.
The reason why $4.0r_{\rm g}$ is the resonance radius is that the
resonance condition for horizontal resonances of g-mode oscillations
is $2\kappa=\Omega$ (see section 5), and this is realized there in 
the case of the Schwarzschild metric.
The frequencies of the resonant oscillations are $\kappa$,
$\Omega\pm\kappa$, $2\Omega\pm \kappa$,... at $r=4r_{\rm g}$ 
(see Paper I), depending on the modes ($m$ and $n$) of the oscillations.
In units of $(GM/r_{\rm g}^3)^{1/2}$, these frequencies are
\begin{equation}
     0.0625, \quad 0.1875, \quad 0.313, \quad 0.4375,
\label{8.1}
\end{equation}
and their ratios are 1 : 3 : 5 : 7...
The modes of oscillations realizing these frequencies 
are summarized in table 2.

If we are concerned with what frequencies are observable in luminosity 
variations, some other frequencies should 
be added in the list of equation (\ref{8.1}) for the following reason.
Let us consider one-armed ($m=1$) oscillations with frequency
$\omega$.
Then, a one-armed pattern revolves around the disk center with
frequency $\omega$.
After half a revolution, however, the arm appears on the another side of 
the disk.
Hence, if we observe luminosity variations (not velocity variations)
resulting from the whole disk, the phases of the luminosity variations 
become the same as the initial one after half a revolution.
That is, we observe a luminosity variation with frequency $2\omega$.
As shown in table 2, the oscillations of $\kappa$ (or $\Omega-\kappa$)
are realized by one-armed oscillations.
Hence, we can expect luminosity variations with frequency $2\kappa (=\Omega)$.
A similar argument shows that we can also expect a luminosity variation of
$2(\Omega+\kappa)$ or $2(2\Omega-\kappa)$, which is $6\kappa$, since 
oscillations with $\Omega+\kappa$ or $2\Omega-\kappa$ occur for one-armed
oscillations.
The above argument, however, cannot be applied to $m=2$ oscillations.
One-armed oscillations have a unique characteristic in this sense.
Summing them, we can say that the observed frequencies are some hamonics of 
$\kappa$, as follows:
\begin{equation}
   0.0625, \quad 0.125, \quad 0.1875, \quad 0.313,..
\label{8.2}
\end{equation}
in the case of the Schwarzschild metric.
Their ratios are 1 : 2 : 3 : 5 : 6 : 7 ...

The lowest frequency of oscillations, $\kappa$, is hidden by the
oscillations of $2\kappa$.
The most observable pairs of oscillations are thus 2 : 3.
This will be the reason why 2 : 3 oscillations are prominently observed as
QPOs in black hole candidates.
The above arguments suggest that other sets of oscillations, for example,
1 : 2 and 3 : 5, are also observable candidates.
It is noted that the importance of 2 : 3 pair oscillations was first   
emphasized by Abramowicz and Klu{\' z}niak (2001) and
Klu{\' z}niak and Abramowicz (2001), although they did not 
present definite resonant mechanisms.

Next, we estimate the masses (and spins) of four black-hole objects  
with pairs of QPOs.
In all of them, the pairs are close to 2 : 3.
If we take the observed upper frequencies of these QPOs to be
$3\kappa(r_{\rm max})$ with the Schwarzschild metric, 
the masses of the central black hole can be calculated for each object.
The results are shown on the fourth line of table 3.
The second line of table 3 shows the observed pairs of frequencies in units
of hertz, and the bottom line of the table shows the mass
ranges (in units of the solar mass) which were evaluated from
the observed data.
In XTE 1550$-$564 and GRO 1655$-$40, the masses estimated by our model are 
slightly below the mass ranges obtained from observations.
One of the reasons would be that effects of the rotation of the 
central objects are not taken into account in evaluating the frequencies. 

The processes for deriving the resonance condition $2\kappa=\Omega$ 
in this paper
suggest that even when the central object has rotation, the condition
for horizontal resonances of g-mode oscillations remains  
$2\kappa=\Omega$ in the lowest order of approximations. 
Using the Kerr metric with the spin parameter 
$a (0\leq a < 1)$, we can calculate, as a function of $a$, the radius where 
$2\kappa=\Omega$ is realized.
Then, $\kappa$ (or $\Omega$) values at the radius are obtained as functions 
of the mass and spin of the central objects.
Comparing the frequencies calculated in this way with observations, 
we can estimate the masses of the central objects as a function of $a$.
The results are also presented in table 3, which shows that the masses
evaluated by our model are in good agreement with observations 
for moderate values of $a$.
Let us denote the upper frequency of the twin pairs by $\nu_{\rm upp}$.
Then, the relation between $\nu_{\rm upp}$ and the mass, $M$, for a
given $a$ is
\begin{equation}
        \nu_{\rm upp}=A\biggr({M\over M_\odot}\biggr)^{-1} {\rm kHz},
\end{equation}
where $A=2.1$, $2.4$, $2.8$, and $3.4$ for $a=0$, $0.2$, $0.4$, and
$0.6$, respectively.
The relation in the case of $a=0.4$ is just the same as that obtained by
McClintock and Remillard (2003) as a fitting formula of observations.

\begin{longtable}{lcll}
  \caption{Frequencies and modes of resonant oscillations.}\label{tab:2}
  \hline\hline
    Frequencies & & Modes of oscillations & \\ 
  \hline
  & in units of $(GM/r_{\rm g}^3)^{1/2}$ & g-mode (arbitrary $n$) & 
                                           p-mode ($n=0$) \\
\hline
\endhead
\hline
\endfoot
\hline
\endlastfoot
  $\kappa$, $\Omega-\kappa$ & \ 0.0625 & $m=0$, $1$ & $m=0$, $1$ \\
  $\Omega+\kappa$, $2\Omega-\kappa$  &  \ 0.1875 &
                     $m=1$, $2$ & $m=1$, $2$ \\ 
  $2\Omega+\kappa$,$3\Omega-\kappa$ & 0.313\ &
                     $m=2$, $3$ & $m=2$, $3$ \\
\end{longtable}

\begin{longtable}{llcccc}
  \caption{Estimated masses of four black hole candidates.}\label{tab:3}
  \hline\hline
  Objects & &  GRS 1915+105$^{(*)}$ & XTE 1550$-$564$^{(*)}$ 
            & GRO 1655$-$40$^{(*)}$ & H 1743$-$322$^{(+)}$   \\
  Frequencies & &  (113, 168)   & (184, 276)   & (300, 450)  & (160, 240)   \\ 
 \hline
\endfirsthead
  $a$ & $r_{\rm c}/r_{\rm g}$ & $M/M_\odot$ & $M/M_\odot$ 
  & $M/M_\odot$ & $M/M_\odot$ \\ 
  \hline
  0.0 & 4.0  & 12.6 &  \ \ \ 8.19 & 4.75 & \ \ \ 8.90 \\
  0.1 & 3.84 & 13.4 &  \ \ \ 8.78 & 5.04 & \ \ \ 9.44 \\
  0.2 & 3.66 & 14.3 &  \ \ \ 9.26 & 5.37 & 10.1  \\
  0.3 & 3.48 & 15.3 &  \ \ \ 9.93 & 5.76 & 10.8  \\
  0.4 & 3.29 & 16.5 & 10.7  & 6.22 & 11.7 \\
  0.5 & 3.08 & 18.0 & 11.7  & 6.79 & 12.7 \\
\hline
      &      & 10.0 -- 18.0 & 8.4 -- 10.8 & 6.0 -- 6.6 & not measured \\
\hline
\end{longtable}
\footnotetext{(*) McClintock \& Remillard (2003),
              (+) Homan et al. (2003). See also Abramowicz et al. (2004b).} 

The frequency ratios of some observed QPO pairs are slightly deviated from
2 : 3.
This occurs natually in our model, since oscillations propagate in the radial 
direction as a wave packet, changing the resonant frequencies and their 
wavelengths.
Another possibility of the deviation is precession of the warp.
A precession of the warp is generally expected when the central object 
has rotation.
If this interpretation is correct, the frequency of the precession
is related to the deviation, and the precession might be
observed as a low-frequency QPO.
Large $Q$-values of QPOs are also expected in our model, since the
oscillations are not trapped standing ones.
Furthermore, different modes of oscillations with different paths of
couplings appear with close frequencies, as shown in table 2
(see also Paper I).

If our present model represents the real mechanism of kHz QPOs, observations
of QPOs are a powerful tool for evaluating the mass and spin of the central 
black holes.
Furthermore, it is implied that a warp occasionally appears in the
innermost region of relativistic disks.
This would be of importance for understanding the activities, structures, and 
evolutions of the innermost region of black-hole disks.

Finally, it is noted that the basic process of our resonance model is
applicable to any deformed disks.
For example, if the disks are deformed by the spin of the central objects 
through a spin-disk coupling, we can expect disk oscillations whose 
frequencies are correlated to the spin frequency.
It is known that in some X-ray binaries with neutron stars,
the frequency difference between the high frequency QPO pairs is equal to, 
or half of, the spin frequency (Wijnands et al. 2003).
In a subsequent paper we will examine whether the phenomenon could be
explained by a simple extension of our present resonance model.

\bigskip\noindent
{\bf Appendix 1. Expressions for $C$ in Cylindrical Coordinates}

The coupling term $\mbox{\boldmath $C$}(\mbox{\boldmath $\xi$},
\mbox{\boldmath $\xi$})$ consists of two terms as 
$\mbox{\boldmath $C$}=\mbox{\boldmath $C$}_\psi +
\mbox{\boldmath $C$}_p$.
Expressions for 
$\mbox{\boldmath $C$}_\psi$ and $\mbox{\boldmath $C$}_p$ in 
cylindrical coordinates are
\begin{equation}
  \rho_0C_{\psi,r}=-{1\over 2}\rho_0\biggr[
     \biggr(\xi_r^2{\partial^2\over\partial r^2}       
     +2\xi_r\xi_z{\partial^2\over\partial r\partial z}
     +\xi_z^2{\partial^2\over\partial z^2}\biggr)
             {\partial\psi_0\over\partial r}
     -{\xi_\varphi^2\over r^2}{\partial\psi_0\over\partial r}
     +{\xi_\varphi^2\over r}{\partial^2\psi_0\over\partial r^2}\biggr],
\end{equation}
\begin{equation}
  \rho_0C_{\psi,\varphi}=-\rho_0\biggr[
        \xi_r\xi_\varphi{\partial\over\partial r}
              \biggr({1\over r}{\partial\psi_0\over\partial r}\biggr)
       +{\xi_\varphi\xi_z\over r}{\partial^2\psi_0\over\partial r\partial z}
       \biggr],
\end{equation}
\begin{equation}
  \rho_0C_\psi,z=-{1\over 2}\rho_0\biggr[
     \biggr(\xi_r^2{\partial^2\over\partial r^2}       
     +2\xi_r\xi_z{\partial^2\over\partial r\partial z}
     +\xi_z^2{\partial^2\over\partial z^2}\biggr)
             {\partial\psi_0\over\partial z}
     +{1\over r}\xi_\varphi^2{\partial^2\psi_0\over\partial r\partial z}
     \biggr],
\end{equation}
\begin{equation}
   \rho_0C_{p,r}=-{\partial T_{rr}\over\partial r}
           -{\partial T_{\varphi r}\over r\partial\varphi}
           -{\partial T_{zr}\over\partial z}
           -{1\over r}T_{rr}+{1\over r}T_{\varphi\varphi},
\end{equation}
\begin{equation}
   \rho_0C_{p,\varphi}=
           -{\partial T_{r\varphi}\over\partial r}
           -{\partial T_{\varphi\varphi}\over r\partial\varphi}
           -{\partial T_{z\varphi}\over\partial z}
           -{1\over r}(T_{r\varphi}+T_{\varphi r}),
\end{equation}
\begin{equation}
   \rho_0C_{p,z}=-{\partial T_{rz}\over\partial r}
           -{\partial T_{\varphi z}\over r\partial\varphi}
           -{\partial T_{zz}\over\partial z}
           -{1\over r}T_{rz},
\end{equation}
where
\begin{equation}
   T_{rr}=p_0\biggr[
           \biggr({\partial\xi_r\over\partial r}\biggr)^2
          +\biggr({\partial\xi_r\over r\partial\varphi}
            -{\xi_\varphi\over r}\biggr){\partial\xi_\varphi\over\partial r}
          +{\partial\xi_r\over\partial z}{\partial\xi_z\over \partial r}
          \biggr],
\end{equation}
\begin{equation}
    T_{r\varphi}=p_0\biggr[
           \biggr({\partial\xi_r\over\partial r}
                  +{\partial\xi_\varphi\over r\partial \varphi}
                  +{\xi_r\over r}\biggr)
           \biggr({\partial \xi_r\over r\partial\varphi}
                  -{\xi_\varphi\over r}\biggr)
          +{\partial\xi_r\over\partial z}{\partial\xi_z\over r\partial\varphi}
          \biggr],
\end{equation}
\begin{equation}
    T_{rz}=p_0\biggr[
          {\partial\xi_r\over\partial r}{\partial\xi_r\over\partial z}
          +\biggr({\partial\xi_r\over r\partial\varphi}
               -{\xi_\varphi\over r}\biggr){\partial\xi_\varphi\over\partial z}
          +{\partial\xi_r\over\partial z}{\partial\xi_z\over\partial z}
          \biggr],
\end{equation}
\begin{equation}
    T_{\varphi r}=p_0\biggr[
          {\partial\xi_\varphi\over\partial r}{\partial\xi_r\over\partial r}
          +\biggr({\partial \xi_\varphi\over r\partial\varphi}+{\xi_r\over r}
                \biggr){\partial\xi_\varphi\over\partial r}
          +{\partial\xi_\varphi\over\partial z}{\partial\xi_z\over\partial r}
          \biggr],
\end{equation}
\begin{equation}
     T_{\varphi\varphi}=p_0\biggr[
          {\partial\xi_\varphi\over\partial r}
              \biggr({\partial\xi_r\over r\partial\varphi}
              -{\xi_\varphi\over r}\biggr)
          +\biggr({\partial\xi_\varphi\over r\partial\varphi}
                 +{\xi_r\over r}\biggr)^2
          +{\partial \xi_\varphi\over\partial z}
                 {\partial\xi_z\over r\partial\varphi}
          \biggr],
\end{equation}
\begin{equation}
      T_{\varphi z}=p_0\biggr[
           {\partial\xi_\varphi\over\partial r}{\partial\xi_r\over\partial z}
           +\biggr({\partial\xi_\varphi\over r\partial\varphi}
                +{\xi_r\over r}\biggr){\partial\xi_\varphi\over\partial z}
           +{\partial\xi_\varphi\over\partial z}{\partial\xi_z\over\partial z}
           \biggr],
\end{equation}
\begin{equation}
     T_{zr}=p_0\biggr[
          {\partial\xi_z\over\partial r}{\partial\xi_r\over\partial r}
          +{\partial\xi_z\over r\partial\varphi}
                      {\partial\xi_\varphi\over\partial r}
          +{\partial\xi_z\over\partial z}{\partial\xi_z\over\partial r}
          \biggr],
\end{equation}
\begin{equation}
     T_{z\varphi}=p_0\biggr[
          {\partial\xi_z\over\partial r}
               \biggr({\partial\xi_r\over r\partial\varphi}
                      -{\xi_\varphi\over r}\biggr)
          +{\partial\xi_z\over r\partial\varphi}
               \biggr({\partial\xi_\varphi\over r\partial\varphi}
                      +{\xi_r\over r}\biggr)
          +{\partial\xi_z\over\partial z}{\partial\xi_z\over r\partial\varphi}
          \biggr],
\end{equation}
and
\begin{equation}
      T_{zz}=p_0\biggr[
          {\partial\xi_z\over \partial r}{\partial\xi_r\over\partial z}
          +{\partial\xi_z\over r\partial\varphi}
                    {\partial\xi_\varphi\over\partial z}
          +{\partial\xi_z\over\partial z}{\partial\xi_z\over\partial z}\biggr].
\end{equation}

\bigskip\noindent
{\bf Appendix 2. Expressions for ${\hat C_{r,\pm,{\tilde n}}^{(1)}}$,
         ${\hat C_{\varphi,\pm,{\tilde n}}^{(1)}}$, and 
         ${\hat C_{z,\pm,{\tilde n}}^{(1)}}$}

From the expressions in appendix 1, we have
\begin{eqnarray}
  {\hat C}_{r,+,n+1}^{(1)}&&=\Omega_\bot^2 H\biggr\{-{1\over \Omega_\bot^2}
         {d^2\Omega_\bot^2\over dr^2} {\hat \xi}_r\eta \nonumber \\
        &&+{d{\hat \xi}_r\over dr}{d\eta\over dr}
          -i{d{\hat \xi}_\varphi \over dr}{\eta\over r}
          +n{d{\rm ln}H\over dr}\biggr[-2{\hat \xi}_r{d\eta\over dr}+
                                  i{\hat \xi}_\varphi{\eta\over r}\biggr]
                             \biggr\},
\label{a2.1}
\end{eqnarray}
\begin{eqnarray}
  {\hat C}_{r,+,n-1}^{(1)}&&=\Omega_\bot^2 H\biggr\{-n{1\over \Omega_\bot^2}
                    {d^2\Omega_\bot^2\over dr^2}{\hat \xi}_r \eta
                  -{d{\ln \Omega}_\bot^2\over dr}{{\hat \xi}_z\over H}\eta
                  -{n\over rp_{00}H}{d\over dr}\biggr(rp_{00}{\hat \xi}_r
                          {d\eta\over dr}\biggr)  \nonumber \\
            &&
         +n(n-1){d{\rm ln}H\over dr}\biggr[ -2{\hat \xi}_r{d\eta\over dr}+
                   i{\hat \xi}_\varphi{\eta\over r}\biggr]
         +in{{\hat \xi}_\varphi \over r}\biggr[(m+1){d\eta\over dr}-
                              {\eta\over r}\biggr]
         +(n-1){{\hat \xi}_z\over H}{d\eta\over dr}\biggr\},
\label{a2.2}
\end{eqnarray}
\begin{eqnarray}
  {\hat C}_{\varphi,+,n+1}^{(1)}&&=\Omega_\bot^2 H \biggr\{
                       -{d{\rm ln}\Omega_\bot^2\over dr}{\hat \xi}_\varphi
                            {\eta\over r}    \nonumber \\
   && -(im{\hat \xi}_r+{\hat \xi}_\varphi){d\eta\over rdr}
      -(m{\hat \xi}_\varphi+i{\hat \xi}_r){\eta\over r^2}
      +in{d{\rm ln}H\over dr}{\hat \xi}_r{\eta\over r} \biggr\},
\label{a2.3}
\end{eqnarray}
\begin{eqnarray}
  {\hat C}_{\varphi,+,n-1}^{(1)}&&=\Omega_\bot^2 H\biggr\{
  -n{d{\rm ln}\Omega_\bot^2\over dr}{\hat \xi}_\varphi {\eta\over r}
  +i{n\over p_{00}H^{n-1}}{d\over dr}\biggr(p_{00}H^{n-1}{\hat \xi}_r
        \eta \biggr)
     \nonumber \\
   &&+{n\over r}\biggr[ (m+1){\hat \xi}_\varphi{\eta\over r}-{\hat \xi}_\varphi
                   {d\eta\over dr}\biggr]
   -i(n-1){{\hat \xi}_z\over H}{\eta\over r}\biggr\},
\label{a2.4}
\end{eqnarray}
\begin{equation}
    {\hat C}_{z,+,n+1}^{(1)}=\Omega_\bot^2\biggr\{
        -{d{\rm ln}{\rm \Omega}_\bot^2\over dr}{\hat\xi}_r\eta
        +9{H\over r^2}{\hat\xi}_z\eta   
        +n\biggr({\hat\xi}_r{d\eta\over dr}-i{\hat\xi}_\varphi{\eta\over r}
         \biggr)\biggr\},
\label{a2.5}
\end{equation}
\begin{equation}
   {\hat C}_{z,+,n-1}^{(1)}=3(n-1)\Omega_\bot^2{H^2\over r}
      \biggr\{3{{\hat\xi}_z\over H}{\eta\over r}+
       n\biggr({d{\rm ln}\Omega_\bot^2\over dr}-2{\Omega_\bot^2\over r}\biggr)
         {\hat\xi}_r{\eta\over r} \biggr\}.
\label{a2.6}
\end{equation}
The above expressions for the ${\hat C}$'s are those for the path of
$m$ $\rightarrow$ $m+1$ $\rightarrow$ $m$ (notice + in
the subscript).
The expressions for the ${\hat C}$'s resulting from the path of 
$m$ $\rightarrow$ $m-1$ $\rightarrow$ $m$ (i.e, ${\hat C}$'s
with the subscript of $-$ ) are similar to those with a + sign.
That is, expressions for ${\hat C}$'s with the - sign are obtained by 
changing $i$ and $m$ in the expressions for ${\hat C}$'s with
the + sign to $-i$ and $-m$, respectively.
Furthermore, $\eta$ is changed to $\eta^*$.

\bigskip\noindent
{\bf Appendix 3. Expressions for $\hat{C}_{r,\pm,{\tilde n}}^{(2)}$,
                     $\hat{C}_{\varphi,\pm,{\tilde n}}^{(2)}$, 
                       and $\hat{C}_{z,\pm,{\tilde n}}^{(2)}$}

Expressions for ${\hat C}_{r,\pm,{\tilde n}}^{(2)}$ can be derived from 
$\mbox{\boldmath $C$}$'s given in appendix 1 by similar procedures
as in appendix 2.
The results of manipulations show that ${\hat C}_{r,+,n+1}^{(2)}
(\mbox{\boldmath $\zeta$}_{+,n+1}^{(1)}, \mbox{\boldmath $\eta$}^*)$ is
related to ${\hat C}_{r,-,n-1}^{(1)}
(\mbox{\boldmath $\xi$}_{-,n-1}^{(0)}, \mbox{\boldmath $\eta$}^*)$.
That is, the former expression is obtained from the latter one by changing 
the argument, $\mbox{\boldmath $\xi$}_{-,n-1}^{(0)}$, in the latter to 
$\mbox{\boldmath $\zeta$}_{+,n+1}^{(1)}$ and futher $m$ and $n$ in the 
latter, respectively, to $m+1$ and $n+1$.
Similarly,  ${\hat C}_{r,+,n-1}^{(2)}
(\mbox{\boldmath $\zeta$}_{+,n-1}^{(1)}, \mbox{\boldmath $\eta$}^*)$ 
is obtained from 
${\hat C}_{r,-,n+1}^{(1)}
(\mbox{\boldmath $\xi$}_{-,n+1}^{(0)}, \mbox{\boldmath $\eta$}^*)$ 
by changing $\mbox{\boldmath $\xi$}_{-,n+1}^{(0)}$ in the latter to
$\mbox{\boldmath $\zeta$}_{+,n-1}^{(1)}$, and $m$ and $n$ to $m+1$ and
$n-1$ , respectively.

The expression for ${\hat C}_{r,-,n+1}^{(2)}
(\mbox{\boldmath $\zeta$}_{-,n+1}^{(1)}, \mbox{\boldmath $\eta$})$ 
is obtained from  ${\hat C}_{r,+,n-1}^{(1)}
(\mbox{\boldmath $\xi$}_{+,n-1}^{(0)}, \mbox{\boldmath $\eta$})$ 
by changing $\mbox{\boldmath $\xi$}_{+,n-1}^{(0)}$
to $\mbox{\boldmath $\zeta$}_{-,n+1}^{(1)}$ and $m$ and $n$ to $m-1$
and $n+1$, respectively.
Similarly, ${\hat C}_{r,-,n-1}^{(2)}
(\mbox{\boldmath $\zeta$}_{-,n-1}^{(1)}, \mbox{\boldmath $\eta$})$ 
is obtained from 
${\hat C}_{r,+,n+1}^{(1)}
(\mbox{\boldmath $\xi$}_{+,n+1}^{(0)}, \mbox{\boldmath $\eta$})$ 
by changing $\mbox{\boldmath $\xi$}_{+,n+1}^{(0)}$ to
$\mbox{\boldmath $\zeta$}_{-,n-1}^{(1)}$ and $m$ and $n$, respectively, to
$m-1$ and $n-1$.

Expressions for ${\hat C}_{\varphi,\pm,{\tilde n}}^{(2)}$ and 
${\hat C}_{z,\pm,{\tilde n}}^{(2)}$ are neglected here, since they are easily 
supposed from the above procedures.

\newpage


\end{document}